\documentclass[12pt]{article}

\usepackage[latin1]{inputenc}
\usepackage{fancyhdr}
\usepackage{indentfirst}
\usepackage{graphicx}
\usepackage{newlfont}

\usepackage{subfigure}

\usepackage{epsfig}
\usepackage{amssymb}
\usepackage{amsmath}
\usepackage{latexsym}
\usepackage{amsthm}
\usepackage{bbold}

\usepackage{xcolor}

\usepackage[export]{adjustbox}

\begin{document}

\title{Microscopic models for the large-scale spread of SARS-CoV-2 virus:
A Statistical Mechanics approach }

\author{Marzia Bisi$^1$ and Silvia Lorenzani$^2$ \thanks{corresponding author}}

\date{\em $^{1}$ Dipartimento di Scienze Matematiche, Fisiche e Informatiche,
Universit\`a di Parma, Parco Area delle Scienze 7/A, 43124 Parma, Italy \\
$^{2}$ Dipartimento di Matematica, Politecnico di Milano, Piazza Leonardo da
Vinci 32, 20133 Milano, Italy}

\maketitle

\begin{abstract}
In this work, we derive a system of Boltzmann-type equations to describe
the spread of SARS-CoV-2 virus at the microscopic scale, that is by modeling
the human-to-human mechanisms of transmission.
To this end, we consider two populations, characterized by specific
distribution functions, made up of individuals without symptoms (population
$1$) and infected people with symptoms (population $2$).
The Boltzmann operators model the interactions between individuals within
the same population and among different populations with a probability
of transition from one to the other due to contagion or, vice versa, to
recovery.
In addition, the influence of innate and adaptive immune systems is
taken into account.
Then, starting from the Boltzmann microscopic description we derive
a set of evolution equations for the size and mean state of each population
considered.
Mathematical properties of such macroscopic equations, as equilibria and
their stability, are investigated and some numerical simulations are performed
in order to analyze the ability of our model to reproduce the characteristic
features of Covid-19.

\end{abstract}



%\maketitle

\section{Introduction} \label{1}

The history of mathematical models in epidemiology begins with the paper
"{\it Essai d'une nouvelle analyse de la mortalit\'e caus\'ee par la petite
v\'erole}" written by Daniel Bernoulli in $1766$ to analyze the mortality
due to smallpox in England \cite{Bern}.
In this paper, Bernoulli derived a model showing that inoculation against
the virus would increase the life expectancy.
Following the work of Bernoulli, Lambert in $1772$ extended the model by
incorporating age-dependent parameters \cite{Lamb}.
However, this kind of approach has not been developed systematically until
$1911$ when the British medical doctor Ronald Ross published the paper
"{\it The prevention of malaria}", considered the starting point of the
modern mathematical epidemiology \cite{Ross}.
For the first time, in this work, a set of differential equations were
derived to study the discrete-time dynamics of mosquito-borne malaria.
Similar epidemic models were later developed by Kermack and McKendrick,
who founded the deterministic compartmental modeling \cite{KK, KK2, KK3}.
These authors proposed the celebrated SIR model according to which the
population is divided into three groups or compartments: the susceptible
(S), who can contract the disease, the infected (I), who have already
contracted the disease and can transmit it, and the recovered (R), who are
healed.
Within this framework, it is assumed that the probability of infection of a
susceptible is proportional to the number of its contacts with infected
individuals.
The resulting mathematical model, based on a deterministic system of
ordinary differential equations, relies on the following hypotheses:
(i) a homogeneous mixing of the contacts; (ii) the conservation of the total
population; (iii) relatively low rates of interaction.
Over the years, SIR-type models have been extended to include age-dependent
infection, mortality and spatial dependence of the epidemic spread 
\cite{SR}, \cite{ZSJ}, \cite{ZZ}.

However, since the classical epidemiological models are derived at a
macroscopic level, they neglect the heterogeneity of disease transmission
due to the microscopic features of the interactions between individuals.
Relying on recent developments of kinetic models for the description of
social and economic phenomena \cite{DT, Tosc}, in the last few years a
number of works have appeared aimed to connect the distribution of social
contacts with the spreading of a disease in a multi-agent system
\cite{PT}.
This issue has been carried out by integrating a classical compartmental
model (typically a SIR-type model) with a statistical part based on
Boltzmann-like equations describing the formation of social contacts.
Such an approach allows one to obtain various sub-classes of macroscopic
epidemiological models characterized by non-linear incidence rates.
Within the same mathematical framework of multi-agent systems, a different
procedure has been considered in \cite{DMLT}, where kinetic evolution equations
have been derived for the distribution functions of the viral load of the
individuals, that change as a consequence of binary interactions or
interactions with a background.
But also in this case a SIR-like compartmental structure has been taken
into account in order to describe the health status of individuals.

In the present paper, following the general setting put forward in
\cite{Delitala, DLDS}, we derive a system of Boltzmann-type evolution
equations for the distribution function $f_i(t, u_i)$ ($i=1,2$) of two
interacting populations, where the index $i=1$ refers to
individuals without symptoms and $i=2$ to infected people with symptoms.
Unlike the kinetic models mentioned above, in our work the microscopic
variable $u_i \in (-\infty, +\infty)$ indicates the level of the infection,
so that the product $f_i(t, u_i) \, du_i$ gives the number of individuals
of the $i$-th population which at the time $t$ are in the elementary state
$[u_i, u_i+du_i]$.
Within the population described by the distribution function $f_1(t, u_1)$
we can distinguish healthy individuals, when $u_1 <0$, and positive
asymptomatic, when $u_1 \geq 0$.
Similarly, the population represented by the distribution function
$f_2(t, u_2)$ is made up of positive symptomatic, when $u_2 <0$, and
hospitalized individuals, when $u_2 \geq 0$.
Our approach does not rely on one of the existing compartmental models,
but starting from a microscopic description of the interactions between
individuals based on the Boltzmann equation, we derive a new class of
macroscopic models, which account for the peculiarities of Covid-19 spread.
The idea behind our derivation is to treat interactions between individuals
in the same way as those between molecules of chemically reacting gas
mixtures.
In particular, the transition from population $1$ (made up of
individuals without symptoms) to population $2$ (made up of infected
individuals with symptoms) is
described by using the same formalism of reversible chemical reactions
in kinetic theory of gases.
The reverse 'chemical reaction' is taken into account when the medical staff
(belonging to population $1$) 'interacts' with positive symptomatic or
with people hospitalized (belonging to population $2$) giving rise to their
recovery.
We also consider the action of the immune system modeled in close
analogy to the interaction of a particle (a single person) with a
background (the immune system), typical of the neutron transport description
in kinetic theory \cite{Cerci}.
The major mathematical difficulty in our model consists in deriving a system
of four macroscopic equations for the evolution of the number densities and
mean states of healthy people, positive asymptomatic, positive symptomatic
and hospitalized persons, starting from a set of two Boltzmann equations
defined at the microscopic level for the distribution functions $f_1$ and
$f_2$.
The closure of such macroscopic system is based on specific choices of
Boltzmann collision kernels and of interaction rules that, as usual in
multi-agent systems involving human beings, also show stochastic effects.

The rest of the paper is organized as follows.
In Section \ref{2} we present the epidemiological model derived at the
microscopic level by setting a system of Boltzmann-like equations.
In Sections \ref{density_ev_eqs} and \ref{mean} we obtain the macrocopic
equations for the evolution of the size and mean state, respectively,
of each population considered, that is healthy individuals, positive
asymptomatic, positive symptomatic and hospitalized individuals.
These macroscopic equations are then qualitatively analyzed.
In particular, the equilibrium states and their stability are investigated
in Sections \ref{equilibrium} and \ref{stability}, respectively.
Section \ref{numsim} presents numerical test cases carried out in order
to analyze the ability of our model to reproduce the characteristic
features of Covid-19.
Some concluding remarks, mainly aimed at highlighting the novelty of our
approach, are included in Section \ref{final}.
Finally, in Appendix A we introduce some preliminaries on the
scattering kernel formulation
of the Boltzmann equation used in this paper, while Appendix B is
devoted to summarize the main features of the classical kinetic approach to
chemically reacting gas mixtures.

\section{Mathematical formulation } \label{2}

We derive a model describing the spread of SARS-CoV-2 virus within the
general mathematical framework of the kinetic theory for chemically
reacting mixtures of gases \cite{RS}.
The system consists of two populations of interacting individuals.
Each population is denoted by the subscript $i$ ($i=1,2$), according to the
following classification:

\begin{equation}
\left\{
\begin{array}{rl}
i=1: & \text{individuals without symptoms} \\
i=2: & \text{infected  individuals with symptoms.}
\end{array} \right.
\end{equation}
Within the same population, each individual is characterized by a microscopic
state, which is a scalar variable $u_i \in (-\infty, +\infty)$
({activity}) \cite{Delitala}, \cite{DLDS}.
Let us introduce the one-particle distribution function: $f_i=f_i(t, u_i)$.
By definition, the product
$$f_i(t, u_i) \, du_i$$
gives the number of individuals of the $i$-th population which at the time
$t$ are in the elementary state $[u_i, u_i+du_i]$.
The variable $u_i$ depends on the intensity level of a certain pathological
state.
We distinguish the following cases:

\begin{equation} \label{2.2}
f_1(t, u_1): \left\{
\begin{array}{rl}
\text{if} \, u_1 <0 & \rightarrow \text{healthy-individuals} \\
\text{if} \, u_1 \geq 0 & \rightarrow \text{positive-asymptomatic}
\end{array} \right.
\end{equation}

\begin{equation} \label{2.3}
f_2(t, u_2): \left\{
\begin{array}{rl}
\text{if} \, u_2 <0 & \rightarrow \text{positive-symptomatic} \\
\text{if} \, u_2 \geq 0 & \rightarrow \text{hospitalized-individuals}.
\end{array} \right.
\end{equation}
Specifically, within the same population, the larger is the value of $u_i$,
the stronger is the infection.

The evolution of the system is determined by microscopic interactions
between pairs of individuals, which modify the probability distribution over
the state variable and/or the size of the population.
The system is homogeneous in space and only binary interactions are taken
into account.
In addition, we model also the action of the immune system described by the
distribution function $\phi_i(v) \; \; (i=1,2)$ over the microscopic
variable $v \in [-M, +M]$, where $M$ is a constant such that $M >>1$.
The immune response to SARS-CoV-2 virus is unpredictable and very different
from person to person.
We distinguish between two natural actions:

\par\noindent
(i) The innate immunity.

\par\noindent
Bacteria or viruses that enter the body can be stopped right away by the
innate immune system.
The effectiveness of this type of action is linked to the possibility that
an individual belonging to population $1$ becomes ill or not.

\par\noindent
(ii) The adaptive immunity.

\par\noindent
The adaptive immune system takes over if the innate immune system is not
able to destroy the germs.
It is slower to respond than the innate immune system, but it identifies the
germs and it is able to 'remember' them.
To the adaptive immunity can be ascribed the recovery of an individual
belonging to population $2$ even in the absence of specific care.
The current knowledge about SARS-CoV-2 infection indicates that the immune
system plays a crucial role in setting the severity of Covid-19.
Appropriate immune responses against SARS-CoV-2 could mitigate the symptoms
of Covid-19 and prevent the occurrence of a severe disease, while excessive
responses trigger pathogenic cell activation increasing the risk of death.

We will consider the following moments of the distribution functions.

(a) The size of the $i$th population at time $t$:

\begin{equation} \label{2.4}
n_i(t)=\int_{-\infty}^{+\infty} f_i (t, u_i) \, du_i, \; \; \; \; i=1,2
\end{equation}
where $n_1=n_1^{HE}+n_1^A$ and $n_2=n_2^S+n_2^{HS}$, with

\begin{equation} \label{2.4a}
n_1^{HE}=\int_{-\infty}^{0} f_1 (t, u_1) \, du_1: \; \; \text{number density
of healthy individuals}
\end{equation}
\begin{equation} \label{2.4b}
n_1^A=\int_{0}^{+\infty} f_1 (t, u_1) \, du_1: \; \; \text{number density
of asymptomatic}
\end{equation}
\begin{equation} \label{2.4c}
n_2^S=\int_{-\infty}^{0} f_2 (t, u_2) \, du_2: \; \; \text{number density
of symptomatic}
\end{equation}
\begin{equation} \label{2.4d}
n_2^{HS}=\int_{0}^{+\infty} f_2 (t, u_2) \, du_2: \; \; \text{number density
of hospitalized individuals}.
\end{equation}
The total number of individuals at time $t$ is given by:

\begin{equation} \label{2.5}
N=\sum_{i=1}^{2} n_i(t).
\end{equation}
Since in this work we are mainly interested in modeling the transmission
mechanism of the disease, we will neglect changes in the social structure
such as the aging process, births, and deaths.
This conservation property implies that $N$ is a constant.

Furthermore, the overall action of the immune system on the individuals
of the $i$th population is given by

\begin{equation} \label{a.4}
\hat{I}_i= \int_{-M}^{+M} \phi_i (v) \, dv, \; \; \; \; i=1,2.
\end{equation}

Higher-order moments provide additional information on the (macroscopic)
description of the system.

(b) Progression of the epidemiological state:

\begin{equation} \label{2.7}
U_i(t)=\int_{-\infty}^{+\infty} u_i \, f_i(t, u_i) \, du_i \; \; \; \; i=1,2.
\end{equation}

The evolution equations for the distribution functions $f_i(t, u_i)$ can be
obtained following the formalism of the Boltzmann equations for chemically
reacting gas mixtures (see Appendix B).
In order to derive these equations, we consider the following hypotheses
on interaction processes:

\par\noindent
(H.1) the medical staff belongs to the population $1$, with
$u_1 \in [-\infty, 0[$, and never becomes positive-asymptomatic or ill
when interacting with sick people belonging to population $2$;

\par\noindent
(H.2) the interactions produce a smooth shift towards higher pathological
states (e.g., a healthy person does not immediately become ill but
positive-asymptomatic);

\par\noindent
(H.3) all positive-symptomatic individuals are in isolation and, therefore,
they can only have interactions with family members (also positive-symptomatic)
 and with the medical staff;

\par\noindent
(H.4) hospitalized-individuals can only have interactions between themselves
and with the medical staff.

\par\noindent
Relying on these assumptions, we describe the interactions within the same
population or between different populations as follows.

\begin{itemize}
\item Interactions within the population $1$.

\par\noindent
(a) Healthy-individuals + healthy-individuals\\
$\longrightarrow$ both individuals remain healthy (population $1$);

\par\noindent
(b) healthy-individuals + positive-asymptomatic: \\
besides the interactions in which individuals do not change category, it may
occur the transition healthy-individual $\longrightarrow$
positive-asymptomatic (population $1$);

\par\noindent
(c) positive-asymptomatic + positive-asymptomatic:\\
it could happen that both remain positive-asymptomatic, or that one undergoes
the transition
positive-asymptomatic $\longrightarrow$  positive-symptomatic (population~$2$).

\item Interactions within the population $2$.

\par\noindent
(d) Positive-symptomatic + positive-symptomatic:\\
one of them could be subject to the change
positive-symptomatic $\longrightarrow$ hospitalized-individual (population $2$);

\par\noindent
(e) hospitalized-individuals + hospitalized-individuals:\\
$\longrightarrow$ both individuals remain hospitalized (population $2$).

\item Interactions between the populations $1$ and $2$.\\
Only the medical staff interacts with positive symptomatic or with
hospitalized people. Besides 'elastic' encounters, where the ill individuals
do not change their category, we have the following interactions giving rise
to recovery:

\par\noindent
(f) Healthy-individual (medical staff) + positive-symptomatic
$\longrightarrow$ healthy-individuals (transition to population $1$);

\par\noindent
(g) healthy-individual (medical staff) + hospitalized-individuals
$\longrightarrow$ healthy-individuals
(transition to population $1$).
\end{itemize}

\par\noindent
Therefore, we shall deal with a 'mixture' of two populations $(1,2)$
which can interact according to the following reversible 'chemical reaction':

\begin{equation} \label{2.9}
1+1 \rightleftarrows 2+1
\end{equation}
In addition to the interactions between individuals, we consider also those
between an individual and the immune system, modeled in close analogy to
the interaction of a particle with a background, typical of the neutron
transport phenomena in kinetic theory \cite{Cerci}.
In particular, we assume that among the individuals in population $1$, only
asymptomatic carriers 'interact' with their innate immune system with the
probability to become positive symptomatic (transition to population $2$).
Furthermore, we assume that people belonging to population $2$ (both
positive symptomatic and hospitalized individuals) interact with their
adaptive immune system with a probability of recovery (transition to
population $1$ among healthy individuals).
While the interactions between individuals are described by the full
nonlinear Boltzmann operator, those between a person and the immune
system give rise to a linear collision operator.

\subsection{Derivation of Boltzmann-type equations } \label{2.1}

In the following, we consider the scattering kernel formulation of the
Boltzmann collision operator (see Appendix A)
for both 'elastic' ($Q_{ik}$)
and 'chemical' interactions ($\overline{Q}_i^{(r)}$, $\overline{L}_i^{(r)}$).
In particular, the operators $Q_{ii}$ ($i=1,2$) describe the interactions
between individuals within the same population $i$ without a transfer to
the other population; the operators $Q_{12}$ and $Q_{21}$ model the
interactions between the medical staff (belonging to population $1$) and
sick people (belonging to population $2$) which do not produce any change
of category, while the operators $\overline{Q}_i^{(r)}$ account for those
interactions between individuals which give rise to a transition from one
population to the other; finally, the operators $\overline{L}_i^{(r)}$
take into account the interactions of persons belonging to population $i$
with the immune system.

Thus, we get for population $1$:

\begin{equation} \label{2.10}
\frac{\partial f_1}{\partial t}(t, u_1)=Q_{11}+Q_{12}+\overline{Q}_1^{(r)}+
\overline{L}_1^{(r)}
\end{equation}
where

\begin{eqnarray} \label{I.1} \nonumber
Q_{11}=&\displaystyle \int_{-\infty}^{+\infty} \int_{-\infty}^{+\infty}
\eta_{11} (u_*, u^*) \,
A_{11}^{(1)} (u_*, u^*; u_1) f_1 (t, u_*) \, f_1 (t, u^*) \, du_* \, du^* \\
&-f_1 (t, u_1) \displaystyle \int_{-\infty}^{+\infty} \eta_{11} (u_1, u^*) \,
f_1 (t, u^*) \, du^*
\end{eqnarray}

\begin{eqnarray} \label{I.2} \nonumber
Q_{12}=&\displaystyle \int_{-\infty}^{+\infty} \int_{-\infty}^{+\infty}
\eta_{12} (u_*, u^*) \,
A_{12}^{(1)} (u_*, u^*; u_1) f_1 (t, u_*) \, f_2 (t, u^*) \, du_* \, du^* \\
&-f_1 (t, u_1) \displaystyle \int_{-\infty}^{+\infty} \eta_{12} (u_1, u^*) \,
f_2 (t, u^*) \, du^*
\end{eqnarray}

\begin{eqnarray} \label{I.3} \nonumber
\overline{Q}_1^{(r)}=& \displaystyle \int_{-\infty}^{+\infty} \int_{-\infty}^{+\infty}
\eta_{11}^{(r)}(u_*, u^*) \,
{\tilde A}_{11}^{(1)} (u_*, u^*; u_1) f_1 (t, u_*) \, f_1 (t, u^*) \,
du_* \, du^* \\ \nonumber
&-2 f_1 (t, u_1) \displaystyle \int_{-\infty}^{+\infty} \eta_{11}^{(r)} (u_1, u^*)
\,f_1 (t, u^*) \, du^* \\ \nonumber
&+2 \displaystyle \int_{-\infty}^{+\infty} du_* \int_{-\infty}^{+\infty}
\eta_{21}^{(r)} (u_*, u^*) \,
{\tilde A}_{21}^{(1)} (u_*, u^*; u_1) f_2 (t, u_*) \, f_1 (t, u^*) \,
du^* \\
&-f_1 (t, u_1) \displaystyle \int_{-\infty}^{+\infty}
\eta_{21}^{(r)} (u_1, u^*) \,
f_2 (t, u^*) \, du^*
\end{eqnarray}

\begin{eqnarray} \label{II.3} \nonumber
\overline{L}_1^{(r)}=&\displaystyle \int_{-\infty}^{+\infty} du_*
\int_{-M}^{+M} \mu_{2}^{(r)} (u_*, v^*) \,
B_{2}^{(1)} (u_*, v^*; u_1) f_2 (t, u_*) \, \phi_2 (v^*) \, dv^* \\
&-f_1 (t, u_1) \displaystyle \int_{-M}^{+M} \mu_{1}^{(r)} (u_1, v^*)
\, \phi_1 (v^*) \, dv^*
\end{eqnarray}

Likewise, we can write for population $2$:

\begin{equation} \label{2.13}
\frac{\partial f_2}{\partial t}(t, u_2)=Q_{22}+Q_{21}+\overline{Q}_2^{(r)}+
\overline{L}_2^{(r)}
\end{equation}
where

\begin{eqnarray} \label{I.4} \nonumber
Q_{22}=&\displaystyle \int_{-\infty}^{+\infty} \int_{-\infty}^{+\infty}
\eta_{22} (u_*, u^*) \,
A_{22}^{(2)} (u_*, u^*; u_2) f_2 (t, u_*) \, f_2 (t, u^*) \, du_* \, du^* \\
&-f_2 (t, u_2) \displaystyle \int_{-\infty}^{+\infty}
\eta_{22} (u_2, u^*) \, f_2 (t, u^*) \, du^*
\end{eqnarray}

\begin{eqnarray} \label{I.5} \nonumber
Q_{21}=&\displaystyle \int_{-\infty}^{+\infty} \int_{-\infty}^{+\infty}
\eta_{21} (u_*, u^*) \,
A_{21}^{(2)} (u_*, u^*; u_2) f_2 (t, u_*) \, f_1 (t, u^*) \, du_* \, du^* \\
&-f_2 (t, u_2) \displaystyle \int_{-\infty}^{+\infty} \eta_{21} (u_2, u^*) \,
f_1 (t, u^*) \, du^*
\end{eqnarray}

\begin{eqnarray} \label{I.6} \nonumber
\overline{Q}_2^{(r)}=&\displaystyle \int_{-\infty}^{+\infty} \int_{-\infty}^{+\infty}
\eta_{11}^{(r)} (u_*, u^*) \,
{\tilde A}_{11}^{(2)} (u_*, u^*; u_2) f_1 (t, u_*) \, f_1 (t, u^*) \, du_*
\, du^* \\
&-f_2 (t, u_2) \displaystyle \int_{-\infty}^{+\infty} \eta_{21}^{(r)} (u_2, u^*) \,
f_1 (t, u^*) \, du^*
\end{eqnarray}

\begin{eqnarray} \label{II.6} \nonumber
\overline{L}_2^{(r)}=&\displaystyle \int_{-\infty}^{+\infty} du_*
\int_{-M}^{+M}
\mu_{1}^{(r)} (u_*, v^*) \,
B_{1}^{(2)} (u_*, v^*; u_2) f_1 (t, u_*) \, \phi_1 (v^*) \,  dv^* \\
&-f_2 (t, u_2) \displaystyle \int_{-M}^{+M} \mu_{2}^{(r)} (u_2, v^*) \,
\phi_2 (v^*) \, dv^*.
\end{eqnarray}

In the above equations, $\eta_{hk}$ (and $\eta_{hk}^{(r)}$)
is called encounter rate and it describes
the rate of interactions (that is, the number of encounters per unit time)
between individuals of the $h$-th population and individuals of the $k$-th
population, while $\mu_h^{(r)}$ refers to the rate of interaction between
individuals of the $h$-th population and the immune system.
To construct the transition rates, we assume that the interactions between
individuals can be modeled in analogy to the intermolecular potentials
acting between the gas molecules.
Since in the framework of kinetic theory of rarefied gases analytical
manipulations can be carried out in closed form for Maxwell molecules
(see Appendix A) \cite{Cerci},
we restrict ourselves to this kind of interaction, which leads
to assume $\eta_{hk}$ constant.
In particular, when the encounters between individuals belonging
to different groups are forbidden, in the framework of our epidemiological
model, the rates $\eta_{hk}$ vanish.
Thus, for the elastic operator $Q_{11}$, since all types of
interactions between
healthy and positive--asymptomatic people are allowed, we set
$$
\eta_{11}(u_*, u^*) = \bar{\eta}_{11} \qquad \quad \forall\,\, u_* \in
\mathbb{R}, \quad u^* \in \mathbb{R}.
$$
In the operator $Q_{22}$, since positive--symptomatic individuals cannot
interact with hospitalized people, we have
$$
\eta_{22}(u_*, u^*) = \left\{
\begin{array}{ll}
\bar{\eta}_{22} \qquad \quad & \text{if}\ (u_* < 0, u^*<0)\
\text{or}\ (u_* > 0, u^*>0) \\
0 & \text{otherwise}
\end{array}
\right.
$$
The interactions between the two populations $1$ and $2$ describe only the
possible contacts between the ill individuals  and the medical staff
(healthy), therefore
$$
\eta_{12}(u_*, u^*) = \left\{
\begin{array}{ll}
\bar{\eta}_{12} \qquad \quad & \text{if}\ u_* < 0  \quad \text{in}\
f_1(t, u_*)
\\
0 & \text{otherwise}
\end{array}
\right.
$$
$$
\eta_{21}(u_*, u^*) = \left\{
\begin{array}{ll}
\bar{\eta}_{21} \qquad \quad & \text{if}\ u^*<0 \quad \text{in}\
f_1(t, u^*) \\
0 & \text{otherwise}
\end{array}
\right.
$$
and moreover, since both rates refer to the same type of interactions,
we assume
$$
\bar{\eta}_{12} = \bar{\eta}_{21}\,.
$$
For the direct transition $1 +1 \rightarrow 2+1$, that can occur when two
positive-asymptomatic individuals interact, we set
$$
\eta_{11}^{(r)}(u_*, u^*) = \left\{
\begin{array}{ll}
\bar{\eta}_{11}^{(r)} \qquad \quad & \text{if}\ u_* > 0\ \text{and}\
u^* > 0\\
0 & \text{otherwise}
\end{array}
\right.
$$
For the reverse interaction $2+1 \rightarrow 1+1$, taking into account that
population $2$ may interact only with the medical staff we have
$$
\eta_{21}^{(r)}(u_*, u^*) = \left\{
\begin{array}{ll}
\bar{\eta}_{21}^{(r)} \qquad \quad & \text{if}\  u^* < 0 \quad \text{in}\
f_1(t, u^*) \\
0 & \text{otherwise}
\end{array}
\right.
$$
The interactions with the immune system involve
the positive-asymptomatic persons in population $1$ and
all the individuals in
population $2$; for this reason we set
$$
\mu_{1}^{(r)} (u_*, v^*) = \left\{
\begin{array}{ll}
\bar{\mu}_{1}^{(r)}\qquad \quad & \text{if}\ u_* > 0  \quad \text{in}\
f_1(t, u_*)\\
0 & \text{otherwise}
\end{array}
\right.
$$
and
$$
\mu_{2}^{(r)} (u_*, v^*) = \bar{\mu}_{2}^{(r)} \qquad \quad \forall\,\,
u_* \in \mathbb{R}, \quad v^* \in [-M,M]
$$

The modification of the state of interacting individuals
is described by the transition probability density,
$A_{hk}^{(i)} (u_*, u^*; u_i)$, of individuals which are shifted into the
$i$-th population with state $u_i$ due to encounters between an individual
of the $h$-th population in the state $u_*$ with an individual of the
$k$-th population in the state $u^*$.
Likewise, $B_{h}^{(i)} (u_*, v^*; u_i)$ represents the transition probability
density of individuals which are shifted into the $i$-th population with state
$u_i$ due to the interaction between an individual of the $h$-th population in
the state $u_*$ with the immune system characterized by the microscopic
state $v^*$.
The products $\eta_{hk} \, A_{hk}^{(i)}$ and $\mu_{h}^{(r)} \, B_{h}^{(i)}$
give the transition rates.

For stochastic models of interaction between individuals, the transition
probability density $A_{hk}^{(i)}$ satisfies the following properties:

\par\noindent
(i)
\begin{equation} \label{2.11}
A_{hk}^{(i)} (u_*, u^*; u_i)=A_{kh}^{(i)} (u^*, u_*; u_i)
\end{equation}
expressing indistinguishability of individuals;

\par\noindent
(ii)
\begin{equation} \label{2.12}
 \displaystyle \int_{-\infty}^{+\infty} du_i \, A_{hk}^{(i)} (u_*, u^*; u_i) =1 \; \; \; \; \;
\forall h,k.
\end{equation}
We assume that also the probability density $B_{h}^{(i)}$ is normalized with
respect to all possible final states:

\begin{equation} \label{a.12}
 \displaystyle \int_{-\infty}^{+\infty} du_i \, B_{h}^{(i)} (u_*, v^*; u_i) =1
\; \; \; \; \;
\forall h.
\end{equation}

Stochastic models, describing the interactions within each population,
have been proposed in order to give an explicit expression for the
transition probabilities $A_{hk}^{(i)}$ and $B_{h}^{(i)}$
in the collisional operators of the
Boltzmann equations (Eqs. (\ref{2.10}) and (\ref{2.13})).

In particular, the interactions within the population $1$, taken into account
by the term $Q_{11}$, can be modeled as follows: if the interacting
individuals are both healthy (with microscopic states $u_*, u^*<0$) or
both positive asymptomatic (with $u_*, u^*>0$), one has

\begin{equation} \label{M.1-eq}
\left\{
\begin{array}{rl}
u'_* & =  u_* \\
u'^* & =  u^*,
\end{array} \right.
\end{equation}
while if a healthy individual (with $u_*<0$) interacts with a positive
asymptomatic (with $u^*>0$) the output is

\begin{equation} \label{M.1-bern}
\left\{
\begin{array}{rl}
u'_* & =  (1 - \theta) u_* + \theta u^* \\
u'^* & =  (1 - \theta) u_* + \theta u^*
\end{array} \right.
\end{equation}
where $\theta$ denotes a Bernoulli random variable taking the value
$\theta = 1$ with probability $\beta \in [0,1]$ and the value $\theta = 0$
with probability $1 - \beta$.
Consequently, the expected value of both post--interaction states $u'_*$ and
$u'^*$ is

\begin{eqnarray} \label{lab} \nonumber
E_\theta [u'_*] &= (u'_*)_{|\theta =0} \text{Prob} (\theta =0) +
(u'_*)_{|\theta =1} \text{Prob} (\theta =1) = (1-\beta) u_* + \beta\, u^*\,
\\
E_\theta [u'^*] &= (u'^*)_{|\theta =0} \text{Prob} (\theta =0) +
(u'^*)_{|\theta =1} \text{Prob} (\theta =1) = (1-\beta) u_* + \beta\, u^*\,
\end{eqnarray}
Thus, the parameter $\beta$ represents the contagious index of the disease:
the limiting option $\beta =1$ describes an extremely contagious disease
(pandemic),
in which each healthy individual interacting with a
positive asymptomatic  becomes
infectious, and a positive asymptomatic individual never recovers in these
interactions; the other limit $\beta=0$ corresponds to a situation in which
a healthy individual cannot become infectious and the positive asymptomatic
persons immediately recover, therefore no pandemic is ongoing.
The intermediate value $\beta=1/2$ provides
$E_\theta [u'_*] = E_\theta [u'^*] = \frac12 (u_* + u^*)$,
therefore the probability for a healthy individual to get sick equals the
probability for an asymptomatic person to recover, and
this implies (as it will be
proven below) that the total number of healthy and positive asymptomatic
individuals does not change.

Concerning the interactions described by the operator $Q_{12}$, only a
doctor (healthy, with $u_*<0$) can interact with people of population $2$
(with state $u^* \in \mathbb{R}$), providing

\begin{equation} \label{M.12}
\left\{
\begin{array}{rl}
u'_* & =  u_* \\
u'^* & =  u^*.
\end{array} \right.
\end{equation}

As regards the rules underlying the transition probabilities in the term
$Q_{22}$, interactions between hospitalized people do not cause changes
($u'_* =  u_*$ and $u'^* =  u^*$), while an interaction between two positive
symptomatic leads to a possible aggravation of the status of one of them,
described through
a Bernoulli variable $\theta$ such that if $\theta=1$ the status does not
change while if $\theta=0$ the status changes sign and the individual becomes
hospitalized. Post--interaction states are thus provided by

\begin{equation} \label{M.2}
\left\{
\begin{array}{rl}
u'_* & =  \theta u_* + (1-\theta) (- u^*)  \\
u'^* & = u^*
\end{array} \right.
\end{equation}
and, denoting with $\alpha \in [0,1]$ the probability that $\theta = 1$,
the expected post--collision state is

$$
E_\theta [u'_*] = (u'_*)_{|\theta =0} \text{Prob} (\theta =0) +
(u'_*)_{|\theta =1} \text{Prob} (\theta =1) = \alpha\, u_* +
(1-\alpha)\, (-\,u^*).
$$

Furthermore, encounters between individuals which generate a population
transition can be modeled as follows:

\par\noindent
(a) interaction rule for the direct transition $1+1 \rightarrow 2+1$ with
states $(u^*>0, u_*>0) \rightarrow (u'^*<0, u'_*>0)$, respectively:

\begin{equation} \label{M.3}
\left\{
\begin{array}{rl}
u'_* & = u_*  \\
u'^* & =- u^* \\
\end{array} \right.
\end{equation}
thus an asymptomatic individual becomes symptomatic;

\par\noindent
(b)  interaction rule for the reverse transition $2+1 \rightarrow 1+1$
with states $(u^* \in \mathbb{R}, u_*<0) \rightarrow (u'^*<0, u'_*<0)$,
respectively, in which an ill individual recovers due to interactions with
medical staff:

\begin{equation} \label{M.3b}
\left\{
\begin{array}{rl}
u'_* & = u_*  \\
u'^* & = u_*
\end{array} \right.
\end{equation}

Finally, we describe the interactions of the individuals with the
immune system as follows:

\par\noindent
(a) interaction rules with the innate immune system described by the
distribution function $\phi_1$, where some asymptomatic individuals
($u_* >0$) pass to population $2$ becoming symptomatic ($u'_*<0$):

\begin{equation} \label{M.6}
\left\{
\begin{array}{rl}
u'_* & =-\, u_*  \\
v'^* & =v^*
\end{array} \right.
\end{equation}

\par\noindent
(b) interaction rules with the adaptive immune system, modeled by the
distribution function $\phi_2$, leading people of population $2$ to recover:

\begin{equation} \label{M.5}
\left\{
\begin{array}{rl}
u'_* & = -|u_*|  \\
v'^* & =v^*
\end{array} \right.
\end{equation}

The average action of the immune system on asymptomatic individuals is
described by

\begin{equation} \label{imm1}
\hat{I}_1= \int_{-M}^{+M} \phi_1 (v) \, dv
\end{equation}
while the action on the ill people is accounted for by

\begin{equation} \label{imm2}
\hat{I}_2= \int_{-M}^{+M} \phi_2 (v) \, dv.
\end{equation}
The term $(\bar{\mu}_{1}^{(r)} \hat{I}_1)$ takes very large values when the
innate immune cells of an initially positive--asymptomatic person (who
has already contracted the virus) triggers an exacerbated inflammatory
response (hyperinflammation) leading to a major complication of Covid-19.
Therefore, in this case, a significant fraction of asymptomatic people
become ill and a transition to population $2$ occurs.
On the contrary, a strong response of the adaptive immune system (that is,
a large value of the term $(\bar{\mu}_{2}^{(r)} \hat{I}_2)$) allows an
individual, belonging to population $2$, to recover without specific
treatments.
In fact, adaptive immunity encompasses a set of specific protective
mechanisms against certain pathogens.
Adaptive immunity can also be acquired through the administration of
vaccines, which in turn increase the anti-viral activity of some innate
immune cells \cite{ZXWX}.
Therefore, the impact of vaccination can be taken into account in our model
by increasing the term $(\bar{\mu}_{2}^{(r)} \hat{I}_2)$ and, at the same
time, decreasing $(\bar{\mu}_{1}^{(r)} \hat{I}_1)$.
Moreover, explicitly modeling the action of the immune system allows us to
account
for individuals of different ages since older people (with a weaker immune
system) get sick more easily with a higher probability of an unfavorable
outcome of the disease.

By taking into account the interaction rules presented above,
the transition probability densities $A_{hk}^{(i)}$ and $B_{h}^{(i)}$
can be explicitly written.

\par\noindent
(1) For the 'elastic' operator $Q_{11}$, we can distinguish between the
following cases:

(i)

\begin{equation} \label{trans.1}
A_{11}^{(1)} (u_* <0, u^* <0; u_1)=\delta (u_1-u_*)
\end{equation}

(ii)

\begin{equation} \label{trans.2}
A_{11}^{(1)} (u_* >0, u^* <0; u_1)=\beta \, \delta (u_1-u_*)+
(1-\beta) \, \delta (u_1-u^*)
\end{equation}
where $\beta \in [0,1]$.
The value assumed by $\beta$ allows us to take into account different
variants of the main virus characterized by different transmission levels.
In fact, as already remarked above,
the value $\beta=1$ indicates a highly-contagious variant since all
healthy-individuals, which interact with positive-asymptomatic persons,
change their category.
On the contrary, when $\beta=0$ there is no spread of the contagious
disease, since one has the recovery of all asymptomatic individuals.

(iii)

\begin{equation} \label{trans.3}
A_{11}^{(1)} (u_* >0, u^* >0; u_1)=\delta (u_1-u_*).
\end{equation}

\par\noindent
(2) For the 'elastic' operator $Q_{12}$, we have:

\begin{equation} \label{trans.4}
A_{12}^{(1)} (u_* <0, u^* \in \mathbb{R}; u_1)=\delta (u_1-u_*)
\end{equation}

with $A_{12}^{(1)}=A_{21}^{(2)}$.

\par\noindent
(3) For the 'elastic' operator $Q_{22}$, we can distinguish between the
following cases:

(i)

\begin{equation} \label{trans.Q22.1}
A_{22}^{(2)} (u_* >0, u^* >0; u_2)=\delta (u_2-u_*)
\end{equation}

(ii)

\begin{equation} \label{trans.Q22}
A_{22}^{(2)} (u_* <0, u^* < 0; u_2)=\alpha \, \delta (u_2-u_*)+
(1-\alpha) \, \delta (u_2+u^*)
\end{equation}
where $\alpha \in [0,1]$.
If $\alpha=1$, the positive symptomatic individuals remain in the same
infectious state, while if $\alpha=0$ they are exposed to a worsening of
their disease and become hospitalized.

\par\noindent
(4) Encounters between individuals which generate a population transition,
described by the 'chemical' operator $\overline{Q}_1^{(r)}$, give rise
to the following transition probability densities:

(i)

\begin{equation} \label{trans.5}
\tilde {A}_{11}^{(1)} (u_* >0, u^* >0; u_1)=\delta (u_1-u_*)
\end{equation}

(ii)

\begin{equation} \label{trans.6}
\tilde {A}_{21}^{(1)} (u_* \in \mathbb{R}, u^* <0; u_1)=\delta (u_1-u^*)
\end{equation}
while encounters described by the 'chemical' operator $\overline{Q}_2^{(r)}$
give:

(iii)

\begin{equation} \label{trans.6.b}
\tilde {A}_{11}^{(2)} (u_* >0, u^* >0; u_2)=\delta (u_2+u_*).
\end{equation}

To conclude, the transition probability density
$B_{1}^{(2)} (u_* >0, v^*; u_2)$
of individuals which are
shifted into population $2$, with the state $u_2 <0$, due to the
interaction between an individual of population $1$, in the state
$u_* >0$, with the innate immune system, appearing in the term
$\overline {L}_2^{(r)}$, is given by

\begin{equation} \label{trans.8}
B_{1}^{(2)} (u_* >0, v^*; u_2)=\delta (u_2+u_*).
\end{equation}
Furthermore, the transition probability density
$B_{2}^{(1)} (u_* \in \mathbb{R}, v^*; u_1)$
of individuals which are shifted
into population $1$, with the state $u_1 <0$, due to the interaction
between an individual of population $2$ in the state
$u_* \in \mathbb{R}$ with the adaptive immune system,
appearing in the term $\overline {L}_1^{(r)}$, is given by

\begin{equation} \label{trans.9}
B_{2}^{(1)} (u_* \in \mathbb{R}, v^*; u_1)=\delta (u_1+|u_*|).
\end{equation}

Since this last interaction allows a complete recovery of positive-symptomatic
or hospitalized individuals ($u_1 \in [-\infty, 0[$), the following
relation holds:

\begin{equation} \label{trans.7}
 \displaystyle \int_{-\infty}^{0} du_1 \, B_{2}^{(1)} (u_*, v^*; u_1) =1.
\end{equation}

Global existence and uniqueness of the solutions to the Boltzmann-type
equations, based on the scattering kernel formulation of the collision
operator, have been proven in \cite{DLDS}.

\section{Evolution equations for the size of each population }
\label{density_ev_eqs}

In order to derive the macroscopic equations for the evolution of the
size of the two populations, we have to integrate Eq. (\ref{2.10}) with
respect to $u_1$
and Eq. (\ref{2.13}) with respect to $u_2$.
Taking into account that 'elastic' transitions do not give any contribution
(since they do not change the number of individuals of populations 1 and 2) we
get

\begin{equation} \label{pop1}
\frac{d n_1}{d t}= \int_{-\infty}^{+\infty}
\overline{Q}_1^{(r)}(u_1)\, du_1 + \int_{-\infty}^{+\infty}
\overline{L}_1^{(r)} (u_1)\, du_1\,,
\end{equation}
\begin{equation} \label{pop2}
\frac{d n_2}{d t}= \int_{-\infty}^{+\infty}
\overline{Q}_2^{(r)}(u_2)\, du_2 + \int_{-\infty}^{+\infty}
\overline{L}_2^{(r)} (u_2)\, du_2\,.
\end{equation}
Bearing in mind the definitions of the interaction rates $\eta_{hk}^{(r)}$,
${\mu}_h^{(r)}$
and the transition probability densities $A_{hk}^{(i)}$, $B_{h}^{(i)}$
given above, the equations read:

\begin{equation} \label{dens.1}
\frac{d n_1}{d t}= -\, \bar{\eta}_{11}^{(r)}\, (n_1^A)^2 +
\bar{\eta}_{21}^{(r)}\, n_2\,n_1^{HE} + \bar{\mu}_2^{(r)} \, n_2 \, \hat{I}_2
- \bar{\mu}_1^{(r)} \, n_1^A \, \hat{I}_1
\end{equation}
\begin{equation}  \label{dens.0}
\frac{d n_2}{d t}= \bar{\eta}_{11}^{(r)}\, (n_1^A)^2 -
\bar{\eta}_{21}^{(r)}\, n_2\, n_1^{HE} - \bar{\mu}_2^{(r)} \, n_2 \, \hat{I}_2
+ \bar{\mu}_1^{(r)} \, n_1^A \, \hat{I}_1
\end{equation}
In order to find a closed system of equations, let us rewrite the terms
on the right-hand side of Eq. (\ref{2.10}) as follows, taking into account
the interaction rules (\ref{trans.1})-(\ref{trans.9}).

\begin{eqnarray} \label{dens.2} \nonumber
&Q_{11}=\overline{\eta}_{11}  \displaystyle \int_{-\infty}^{+\infty} du_*
\, f_1(t,u_*) \bigg\{ \displaystyle \int_{-\infty}^{0}
A_{11}^{(1)} (u_*, u^*; u_1) \, f_1(t,u^*) \, du^* \\ \nonumber
&+  \displaystyle
\int_{0}^{+\infty} A_{11}^{(1)} (u_*, u^*; u_1) \, f_1(t,u^*) \, du^*
\bigg \}
-\overline{\eta}_{11} \, f_1(t,u_1) \bigg\{ \displaystyle \int_{-\infty}^{0}
f_1(t,u^*) \, du^*+ \displaystyle \int_{0}^{+\infty} f_1(t,u^*) \, du^*
\bigg \} \\ \nonumber
&=\overline{\eta}_{11} \bigg\{ \displaystyle \int_{-\infty}^{0}
 du_* \, f_1(t,u_*)+\displaystyle \int_{0}^{+\infty} du_* \, f_1(t,u_*)
\bigg \} \cdot \displaystyle \int_{-\infty}^{0}
A_{11}^{(1)} (u_*, u^*; u_1) \, f_1(t,u^*) \, du^* \\ \nonumber
&+\overline{\eta}_{11} \bigg\{ \displaystyle \int_{-\infty}^{0}
du_* \, f_1(t,u_*)+\displaystyle \int_{0}^{+\infty} du_* \, f_1(t,u_*)
\bigg \} \cdot \displaystyle \int_{0}^{+\infty}
A_{11}^{(1)} (u_*, u^*; u_1) \, f_1(t,u^*) \, du^* \\ \nonumber
&-\overline{\eta}_{11} f_1(t, u_1) \displaystyle \int_{-\infty}^{0}
du^* \, f_1(t,u^*)-\overline{\eta}_{11} f_1(t, u_1) \displaystyle
\int_{0}^{+\infty} du^* \, f_1(t,u^*) \\ \nonumber
&=\overline{\eta}_{11} \displaystyle \int_{-\infty}^{0} du_* \, f_1(t,u_*)
\displaystyle \int_{-\infty}^{0} A_{11}^{(1)} (u_*, u^*; u_1) \,
f_1(t,u^*) \, du^*  \\ \nonumber
&+2 \, \overline{\eta}_{11} \displaystyle \int_{0}^{+\infty} du_* \, f_1(t,u_*)
\displaystyle \int_{-\infty}^{0} A_{11}^{(1)} (u_*, u^*; u_1) \,
f_1(t,u^*) \, du^* \\ \nonumber
&+\overline{\eta}_{11} \displaystyle \int_{0}^{+\infty} du_* \, f_1(t,u_*)
\displaystyle \int_{0}^{+\infty} A_{11}^{(1)} (u_*, u^*; u_1) \,
f_1(t,u^*) \, du^* \\ \nonumber
&-\overline{\eta}_{11} f_1(t,u_1) \displaystyle \int_{-\infty}^{0}
f_1(t,u^*) \, du^*-\overline{\eta}_{11} f_1(t,u_1) \displaystyle
\int_{0}^{+\infty} du^* \, f_1(t,u^*) \\ \nonumber
&=\overline{\eta}_{11} f_1(t, u_1 <0) \displaystyle \int_{-\infty}^{0}
du^* \, f_1(t,u^*)+2 \overline{\eta}_{11} \, \beta \, f_1(t, u_1 >0)
\displaystyle \int_{-\infty}^{0} du^* \, f_1(t,u^*) \\ \nonumber
&+2 \overline{\eta}_{11} \, (1-\beta) \, f_1(t, u_1 <0)
\displaystyle \int_{0}^{+\infty} du_* \, f_1(t,u_*)+ \overline{\eta}_{11}
f_1(t, u_1 >0) \displaystyle \int_{0}^{+\infty} du^* \, f_1(t,u^*) \\
&-\overline{\eta}_{11} f_1(t, u_1) \displaystyle \int_{-\infty}^{0}
du^* \, f_1(t,u^*)-\overline{\eta}_{11} f_1(t, u_1) \displaystyle
\displaystyle  \int_{0}^{+\infty} du^* \, f_1(t,u^*)
\end{eqnarray}

\begin{eqnarray} \label{dens.3} \nonumber
&Q_{12}=\overline{\eta}_{12} \displaystyle \int_{-\infty}^{0}
du_* \, f_1(t,u_*) \displaystyle \int_{-\infty}^{+\infty}
A_{12}^{(1)} (u_*, u^*; u_1) \,  f_2(t,u^*) \, du^* \\ \nonumber
&-\overline{\eta}_{12} \, f_1(t, u_1<0) \displaystyle \int_{-\infty}^{+\infty}
f_2(t,u^*) \, du^* \\
&=\overline{\eta}_{12} \, f_1(t, u_1<0) \displaystyle \int_{-\infty}^{+\infty}
f_2(t,u^*) \, du^*-\overline{\eta}_{12} \, f_1(t, u_1<0)
\displaystyle \int_{-\infty}^{+\infty} f_2(t,u^*) \, du^*=0
\end{eqnarray}

\begin{eqnarray} \label{dens.4} \nonumber
&\overline{Q}_1^{(r)}=\overline{\eta}_{11}^{(r)}
\displaystyle \int_{0}^{+\infty} \displaystyle \int_{0}^{+\infty}
\tilde{A}_{11}^{(1)} (u_*, u^*; u_1) \, f_1(t,u_*) \, f_1(t,u^*) \, du_* \,
du^* \\ \nonumber
&-2 \overline{\eta}_{11}^{(r)} \, f_1(t, u_1>0) \displaystyle
\int_{0}^{+\infty} f_1(t,u^*) \, du^*\\ \nonumber
&+ 2 \overline{\eta}_{21}^{(r)} \,
\displaystyle \int_{-\infty}^{0} du^* \, f_1(t,u^*) \displaystyle
\int_{-\infty}^{+\infty} \tilde{A}_{21}^{(1)} (u_*, u^*; u_1) \,
f_2(t,u_*) \, du_* \\ \nonumber
&-\overline{\eta}_{21}^{(r)} \, f_1(t, u_1 <0) \displaystyle
\int_{-\infty}^{+\infty} f_2(t,u^*) \, du^* \\ \nonumber
&=- \overline{\eta}_{11}^{(r)} \,
f_1(t, u_1>0) \displaystyle \int_{0}^{+\infty} f_1(t,u^*) \, du^*
+ \overline{\eta}_{21}^{(r)} \, f_1(t, u_1 <0) \displaystyle
\int_{-\infty}^{+\infty} f_2(t,u_*) \, du_*
\end{eqnarray}

\begin{eqnarray} \label{dens.5} \nonumber
&\overline{L}_1^{(r)}=\overline{\mu}_2^{(r)} \displaystyle
\int_{-\infty}^{+\infty} du_* \displaystyle \int_{-M}^{+M}
B_2^{(1)}(u_*, v^*; u_1) \, f_2(t,u_*) \, \phi_2(v^*) \, dv^* \\
&-\overline{\mu}_1^{(r)} \, f_1(t, u_1 >0) \displaystyle
\int_{-M}^{+M} \phi_1(v^*) \, dv^*
\end{eqnarray}
Integrating Eq. (\ref{2.10}) first with respect to $u_1 \in (-\infty, 0)$ and
then with respect to $u_1 \in (0, +\infty)$, one obtains

\begin{equation} \label{dens.6}
\frac{d n_1^{HE}}{d t }=
(1- 2\,\beta) \, \overline{\eta}_{11}\, n_1^{HE} \, n_1^{A} +
\overline{\eta}_{21}^{(r)} \, n_1^{HE} \, n_2+\overline{\mu}_2^{(r)} \,
\hat{I}_2 \, n_2
\end{equation}

\begin{equation} \label{dens.7}
\frac{d n_1^{A}}{d t }=
 (2 \, \beta - 1) \overline{\eta}_{11}\,n_1^{HE} \, n_1^{A}
-\overline{\eta}_{11}^{(r)} \,
(n_1^{A})^2-\overline{\mu}_1^{(r)} \, \hat{I}_1 \, n_1^{A}
\end{equation}
The expressions (\ref{dens.6}), (\ref{dens.7}) and (\ref{dens.0}) form
a closed system of equations that can be solved to give the time evolution
of the size of the populations $n_1^{HE}$, $n_1^{A}$, $n_2$.

In order to investigate separately the time evolution of $n_2^S$ and
$n_2^{HS}$,
analogous computations as those reported above
can be performed starting from Eq. (\ref{2.13}).
Rewriting the terms on the right-hand side as follows:

\begin{eqnarray} \label{dens.2a}  \nonumber
&Q_{22}=\overline{\eta}_{22} \displaystyle \int_{-\infty}^{0}
\displaystyle \int_{-\infty}^{0} A_{22}^{(2)} (u_*, u^*; u_2) \,
f_2(t,u_*) \, f_2(t,u^*) \, du_* \,du^*  \\ \nonumber
&+\overline{\eta}_{22} \displaystyle \int_{0}^{+\infty}
\displaystyle \int_{0}^{+\infty} A_{22}^{(2)} (u_*, u^*; u_2) \,
f_2(t,u_*) \, f_2(t,u^*) \, du_* \,du^*  \\ \nonumber
&-\overline{\eta}_{22} \, f_2(t, u_2 <0) \displaystyle \int_{-\infty}^{0}
f_2(t,u^*) \, du^*-\overline{\eta}_{22} \, f_2(t, u_2 >0)
\displaystyle \int_{0}^{+\infty} f_2(t,u^*) \, du^* \\ \nonumber
&=\overline{\eta}_{22} \, \alpha \, f_2(t, u_2 <0)
\displaystyle \int_{-\infty}^{0} f_2(t,u^*) \,du^* \\ \nonumber
&+\overline{\eta}_{22} \, (1-\alpha) \displaystyle \int_{-\infty}^{0}
\displaystyle \int_{-\infty}^{0} \delta (u_2+u_*) \, f_2(t,u_*) \, f_2(t,u^*)
\, du_* \,du^* \\
&-\overline{\eta}_{22} \, f_2(t, u_2 <0) \displaystyle \int_{-\infty}^{0}
f_2(t,u^*) \,du^*
\end{eqnarray}

\begin{eqnarray} \label{dens.3a} \nonumber
&Q_{21}=\overline{\eta}_{21} \displaystyle \int_{-\infty}^{0}
du^* \, f_1(t,u^*) \displaystyle \int_{-\infty}^{+\infty}
A_{21}^{(2)} (u_*, u^*; u_2) \, f_2(t,u_*) \, du_* \\ \nonumber
&-\overline{\eta}_{21} \, f_2(t,u_2) \displaystyle \int_{-\infty}^{0}
f_1(t,u^*) \, du^* \\
&=\overline{\eta}_{21} \, f_2(t,u_2) \displaystyle \int_{-\infty}^{0}
f_1(t,u^*) \, du^*-
\overline{\eta}_{21} \, f_2(t,u_2) \displaystyle \int_{-\infty}^{0}
f_1(t,u^*) \, du^*=0
\end{eqnarray}

\begin{eqnarray} \label{dens.4a} \nonumber
&\overline{Q}_2^{(r)}=\overline{\eta}_{11}^{(r)}
\displaystyle \int_{0}^{+\infty} \displaystyle \int_{0}^{+\infty}
\tilde{A}_{11}^{(2)} (u_*, u^*; u_2) \, f_1(t,u_*) \, f_1(t,u^*) \, du_* \,
du^* \\ \nonumber
&-\overline{\eta}_{21}^{(r)} \, f_2(t,u_2) \displaystyle \int_{-\infty}^{0}
f_1(t,u^*) \, du^* \\ \nonumber
&=\overline{\eta}_{11}^{(r)} \displaystyle \int_{0}^{+\infty}
\displaystyle \int_{0}^{+\infty} \delta (u_2+u_*) \, f_1(t,u_*) \,
f_1(t,u^*) \, du_* \, du^* \\
&-\overline{\eta}_{21}^{(r)} \, f_2(t,u_2) \displaystyle \int_{-\infty}^{0}
f_1(t,u^*) \, du^*
\end{eqnarray}

\begin{eqnarray} \label{dens.5a} \nonumber
&\overline{L}_2^{(r)}=\overline{\mu}_1^{(r)}
\displaystyle \int_{0}^{+\infty} du_* \displaystyle \int_{-M}^{+M}
B_1^{(2)}(u_*, v^*; u_2) \, f_1(t,u_*) \, \phi_1(v^*) \, dv^* \\
&-\overline{\mu}_2^{(r)} \, f_2(t,u_2) \displaystyle \int_{-M}^{+M}
\phi_2(v^*) \, dv^*
\end{eqnarray}
and integrating Eq. (\ref{2.13}) first with respect to $u_2 \in (-\infty, 0)$
and then with respect to $u_2 \in (0, +\infty)$, we get

\begin{equation} \label{dens.8}
\frac{d n_2^S}{d t }=
- (1- \alpha) \, \overline{\eta}_{22}\, (n_2^S)^2 +
\overline{\eta}_{11}^{(r)} \, (n_1^{A})^2 - \overline{\eta}_{21}^{(r)}
\, n_1^{HE}\, n_2^S
- \overline{\mu}_2^{(r)}\, \hat{I}_2 \, n_2^S + \overline{\mu}_1^{(r)}\,
\hat{I}_1 \, n_1^A
\end{equation}

\begin{equation} \label{dens.9}
\frac{d n_2^{HS}}{d t }=
(1- \alpha) \, \overline{\eta}_{22}\, (n_2^S)^2 - \overline{\eta}_{21}^{(r)}
\, n_1^{HE}\, n_2^{HS} - \overline{\mu}_2^{(r)}\, \hat{I}_2 \, n_2^{HS}
\end{equation}
The sum of equations (\ref{dens.8}) and (\ref{dens.9}) correctly reproduces
Eq. (\ref{dens.0}) for the total density $n_2$.

Usually, in epidemiological models, the so-called basic reproduction number
($R_0$), which describes the dynamics of the infectious class, is defined
in order to quantify the contagiousness or transmissibility of a virus.
The larger the value of $R_0$, the harder it is to control the
spread of an infectious disease.
$R_0$ is rarely measured directly and its values, deduced from mathematical
models, depend critically on the model structure, the initial conditions,
and several other modeling assumptions \cite{HETH}.
In the framework of the present analysis, we propose the following definition
of time-dependent effective reproduction number:

\begin{equation} \label{r0}
R_0(t)=\frac{ \displaystyle \overline{\eta}_{11}^{(r)} \, (n_1^{A})^2+
\overline{\mu}_1^{(r)} \, n_1^A \, \hat{I}_1 }
{\displaystyle  \overline{\eta}_{21}^{(r)} \, n_2 \, n_1^{HE}+
\overline{\mu}_2^{(r)} \, n_2 \, \hat{I}_2 }
\end{equation}
which characterizes the change in the number of the infected individuals
belonging to population $2$ (see Eq. (\ref{dens.0})) \cite{ABBD}.
In Section \ref{numsim}, we assess the reliability of such a definition
(\ref{r0}) as control parameter of the epidemic.

\section{Evolution equations for the mean state of each population}
\label{mean}

We derive from the Boltzmann model presented in Section \ref{2}
the evolution equations for the total values of the internal states

$$
\begin{array}{c}
\displaystyle U_1^{HE} = \int_{-\infty}^0 u_1\, f_1(t,u_1)\, du_1,
\qquad \quad
U_1^A = \int_0^{+\infty} u_1\, f_1(t,u_1)\, du_1, \vspace*{0.2 cm}\\
\displaystyle U_2^S = \int_{-\infty}^0 u_2\, f_2(t,u_2)\, du_2,
\qquad \quad
U_2^{HS} = \int_0^{+\infty} u_2\, f_2(t,u_2)\, du_2
\end{array}
$$
where the superscripts have the same meaning as those appearing in
formulas (\ref{2.4a})-(\ref{2.4d}).
The equation for $U_1^{HE}$ is provided by

\begin{equation}
\frac{d U_1^{HE}}{d t} = \int_{-\infty}^0 u_1 \Big[ Q_{11} + Q_{12} +
\overline{Q}_1^{(r)} + \overline{L}_1^{(r)} \Big] du_1\,,
\end{equation}
and since

$$
\begin{array}{c}
\displaystyle \int_{-\infty}^0 u_1\, Q_{11}\, du_1 = -\,
(2\,\beta-1) \bar{\eta}_{11} U_1^{HE}\, n_1^A, \qquad \qquad
\int_{-\infty}^0 u_1\, Q_{12}\, du_1 =0, \vspace*{0.2 cm}\\
\displaystyle \int_{-\infty}^0 u_1\, \overline{Q}_1^{(r)}\,
du_1 = \bar{\eta}_{21}^{(r)} U_1^{HE} n_2, \qquad \qquad
\int_{-\infty}^0 u_1\, \overline{L}_1^{(r)}\, du_1 =
\bar{\mu}_2^{(r)} \hat{I}_2 (U_2^S - U_2^{HS}),
\end{array}
$$
it turns out to be

\begin{equation} \label{eqU1HE}
\frac{d U_1^{HE}}{d t} = -\,  (2\,\beta-1) \bar{\eta}_{11}
U_1^{HE}\, n_1^A+ \bar{\eta}_{21}^{(r)} U_1^{HE}\, n_2 +
\bar{\mu}_2^{(r)}\, \hat{I}_2\, ( U_2^S - U_2^{HS}).
\end{equation}

Analogously, the equation for $U_1^A$ reads

\begin{equation}
\frac{d U_1^A}{d t} = \int_0^{+\infty} u_1 \Big[ Q_{11} + Q_{12} +
\overline{Q}_1^{(r)} + \overline{L}_1^{(r)} \Big] du_1\,,
\end{equation}
and since

$$
\begin{array}{c}
\displaystyle \int_0^{+\infty} u_1\, Q_{11}\, du_1 =
(2\,\beta-1) \bar{\eta}_{11} U_1^A n_1^{HE}, \qquad \qquad \int_0^{+\infty}
u_1\, Q_{12}\, du_1 = 0, \vspace*{0.2 cm}\\
\displaystyle \int_0^{+\infty} u_1\, \overline{Q}_1^{(r)}\, du_1 =
-\, \bar{\eta}_{11}^{(r)} U_1^A n_1^A, \qquad \qquad \int_0^{+\infty} u_1\,
\overline{L}_1^{(r)}\, du_1 = -\, \bar{\mu}_1^{(r)} U_1^A \hat{I}_1,
\end{array}
$$
we get

\begin{equation} \label{eqU1A}
\frac{d U_1^A}{d t} = (2\,\beta-1) \bar{\eta}_{11} U_1^A n_1^{HE} -
\bar{\eta}_{11}^{(r)} U_1^A n_1^A - \bar{\mu}_1^{(r)} U_1^A \hat{I}_1\,.
\end{equation}

On the same ground we can derive the evolution equations for
$U_2^S$ and $U_2^{HS}$.
Taking into account the following expressions:

$$
\begin{array}{c}
\displaystyle \int_{-\infty}^0 u_2\, Q_{22}\, du_2 =
-(1-\alpha) \bar{\eta}_{22} U_2^{S}\, n_2^S, \qquad \qquad
\int_{-\infty}^0 u_2\, Q_{21}\, du_2 =0, \vspace*{0.2 cm}\\
\displaystyle \int_{-\infty}^0 u_2\, \overline{Q}_2^{(r)}\,
du_2 = -\bar{\eta}_{11}^{(r)} U_1^{A} n_1^A-\bar{\eta}_{21}^{(r)} U_2^{S}
n_1^{HE}, \; \;
\int_{-\infty}^0 u_2\, \overline{L}_2^{(r)}\, du_2 =
-\bar{\mu}_1^{(r)} U_1^A \hat {I}_1-\bar{\mu}_2^{(r)} U_2^S
\hat {I}_2,
\end{array}
$$
and

$$
\begin{array}{c}
\displaystyle \int_0^{+\infty} u_2\, Q_{22}\, du_2 =
-(1-\alpha) \bar{\eta}_{22} U_2^S n_2^{S}, \qquad \qquad \int_0^{+\infty}
u_2\, Q_{21}\, du_2 = 0, \vspace*{0.2 cm}\\
\displaystyle \int_0^{+\infty} u_2\, \overline{Q}_2^{(r)}\, du_2 =
-\, \bar{\eta}_{21}^{(r)} U_2^{HS} n_1^{HE}, \qquad \qquad
\int_0^{+\infty} u_2\,
\overline{L}_2^{(r)}\, du_2 = -\, \bar{\mu}_2^{(r)} U_2^{HS} \hat{I}_2
\end{array}
$$
we obtain

\begin{equation} \label{eqU2S}
\frac{d U_2^S}{d t} = -\, (1 - \alpha)\, \bar{\eta}_{22} U_2^S\, n_2^S -
\bar{\eta}_{11}^{(r)} U_1^A\, n_1^A - \bar{\eta}_{21}^{(r)} U_2^S\, n_1^{HE}
- \bar{\mu}_1^{(r)}\, \hat{I}_1\, U_1^A - \bar{\mu}_2^{(r)}\, \hat{I}_2\, U_2^S
\end{equation}

\begin{equation} \label{eqU2HS}
\frac{d U_2^{HS}}{d t} = -\, (1 - \alpha)\, \bar{\eta}_{22} U_2^S\, n_2^S -
\bar{\eta}_{21}^{(r)} U_2^{HS}\, n_1^{HE}
- \bar{\mu}_2^{(r)}\, \hat{I}_2\, U_2^{HS}
\end{equation}
We write down now the evolution equations for the mean state of each
population, defined as

$$
\hat{U}_1^{HE}= \frac{U_1^{HE}}{n_1^{HE}}\,, \qquad \quad \hat{U}_1^A =
\frac{U_1^A}{n_1^A}\,, \qquad \quad \hat{U}_2^S = \frac{U_2^S}{n_2^S}\,,
\qquad \quad \hat{U}_2^{HS} = \frac{U_2^{HS}}{n_2^{HS}}\,.
$$
Combining the evolution equations for the number densities
(\ref{dens.6}), (\ref{dens.7}), (\ref{dens.8}), (\ref{dens.9})
and those for the total internal states (\ref{eqU1HE}), (\ref{eqU1A}),
(\ref{eqU2S}), (\ref{eqU2HS}), one has

\begin{equation} \label{eqU1HEmedio}
\frac{d \hat{U}_1^{HE}}{d t} = \frac{1}{n_1^{HE}}\, \frac{d U_1^{HE}}{d t} -
\frac{U_1^{HE}}{(n_1^{HE})^2} \frac{d n_1^{HE}}{dt} =
\frac{\bar{\mu}_2^{(r)}\, \hat{I}_2} {n_1^{HE}}\, \Big( n_2^S \, \hat{U}_2^S -
n_2^{HS} \,\hat{U}_2^{HS} -n_2 \, \hat{U}_1^{HE} \Big)
\end{equation}
\begin{equation} \label{eqU1Amedio}
\frac{d \hat{U}_1^A}{d t} = \frac{1}{n_1^A}\, \frac{d U_1^A}{d t} -
\frac{U_1^A}{(n_1^A)^2} \frac{d n_1^A}{dt} = 0
\end{equation}
\begin{equation} \label{eqU2Smedio}
\frac{d \hat{U}_2^S}{d t} = \frac{1}{n_2^S}\, \frac{d U_2^S}{d t} -
\frac{U_2^S}{(n_2^S)^2} \frac{d n_2^S}{dt} =
-\, \frac{n_1^A}{n_2^S} \Big( \bar{\eta}_{11}^{(r)}\, n_1^A +
\bar{\mu}_1^{(r)}\, \hat{I}_1 \Big) \Big( \hat{U}_1^A + \hat{U}_2^S \Big)
\end{equation}
\begin{equation} \label{eqU2HSmedio}
\frac{d \hat{U}_2^{HS}}{d t} = \frac{1}{n_2^{HS}}\, \frac{d U_2^{HS}}{d t} -
\frac{U_2^{HS}}{(n_2^{HS})^2} \frac{d n_2^{HS}}{dt} =
-\, (1- \alpha)\, \bar{\eta}_{22}\, \frac{(n_2^S)^2}{n_2^{HS}}
\Big( \hat{U}_2^S + \hat{U}_2^{HS} \Big)
\end{equation}

\section{Equilibrium states } \label{equilibrium}

\subsection{Size of the populations } \label{sub_eq1}

Starting from the evolution equations for the size of the populations
(Eqs. (\ref{dens.6}), (\ref{dens.7}), (\ref{dens.8}), (\ref{dens.9})),
the equilibrium states are solutions of
the following system:

\begin{equation} \label{eqden1}
(1-2 \beta) \, \overline{\eta}_{11}\, n_1^{HE} \, n_1^{A}+
\overline{\eta}_{21}^{(r)} \, n_1^{HE} \, n_2+
\overline{\mu}_2^{(r)} \,\hat{I}_2 \, n_2=0
\end{equation}

\begin{equation} \label{eqden2}
(2 \beta-1) \, \overline{\eta}_{11} \, n_1^{HE} \, n_1^{A}-
\overline{\eta}_{11}^{(r)} \, (n_1^{A})^2-\overline{\mu}_1^{(r)} \,
\hat{I}_1 \, n_1^{A}=0
\end{equation}

\begin{equation} \label{eqden3}
-(1- \alpha) \, \overline{\eta}_{22} \, (n_2^S)^2 +
\overline{\eta}_{11}^{(r)} \, (n_1^{A})^2 -\overline{\eta}_{21}^{(r)} \,
n_1^{HE} \, n_2^S-\overline{\mu}_2^{(r)} \, \hat{I}_2 \, n_2^S +
\overline{\mu}_1^{(r)} \, \hat{I}_1 \, n_1^A=0
\end{equation}

\begin{equation} \label{eqden4}
(1- \alpha) \, \overline{\eta}_{22} \, (n_2^S)^2 -
\overline{\eta}_{21}^{(r)} \, n_1^{HE} \, n_2^{HS} - \overline{\mu}_2^{(r)} \,
\hat{I}_2 \, n_2^{HS}=0
\end{equation}
These equations are not independent of each other, therefore one can find
an infinite family of equilibrium solutions.
\par\noindent
(i) The above system admits a trivial solution corresponding to:

\begin{equation} \label{soluzione1}
n_1^{A}=n_2^S=n_2^{HS}=0
\end{equation}
while $n_1^{HE}$ can take any positive value.
Since the total number of individuals $N$ given by (\ref{2.5}) is constant,
this equilibrium solution reads as $n_1^{HE}=N$.
Therefore, it corresponds to the eradication of the pandemic.

\par\noindent
(ii) The unique non-trivial solution of Eqs. (\ref{eqden1})-(\ref{eqden4}),
with $n_1^{HE} > 0, \, n_1^{A} > 0, \, n_2^S  > 0, \, n_2^{HS} > 0,$
is given by (endemic equilibrium):

\begin{equation} \label{soluzione2}
n_1^{A}=(\overline{\eta}_{11}^{(r)})^{-1} [(2 \beta-1) \, \overline{\eta}_{11}
\, n_1^{HE}-\overline{\mu}_1^{(r)} \, \hat{I}_1]
\end{equation}

\begin{eqnarray} \label{soluzione3} \nonumber
&n_2^S=[2 (1-\alpha) \overline{\eta}_{22}]^{-1} \bigg \{
-\overline{\eta}_{21}^{(r)} \, n_1^{HE}-\overline{\mu}_2^{(r)}\, \hat{I}_2+
\bigg [ \bigg ( {\overline{\eta}_{21}^{(r)}}^2+4 (1-\alpha)(2 \beta-1)^2 \,
\frac{\displaystyle \overline{\eta}_{11}^2 \, \overline{\eta}_{22}}
{\displaystyle \overline{\eta}_{11}^{(r)}} \bigg ) (n_1^{HE})^2 \\
&+\bigg (2 \, \overline{\eta}_{21}^{(r)} \, \overline{\mu}_2^{(r)} \,
\hat{I}_2 -4 (1-\alpha) \, (2 \beta-1) \, \frac{\displaystyle
\overline{\eta}_{11} \, \overline{\eta}_{22}}{\displaystyle
\overline{\eta}_{11}^{(r)}} \, \overline{\mu}_1^{(r)}\, \hat{I}_1 \bigg ) \,
n_1^{HE} +{\overline{\mu}_2^{(r)}}^2 \, {\hat{I}_2}^2 \bigg ]^{
\frac{1}{2}} \bigg \}
\end{eqnarray}

\begin{eqnarray} \label{soluzione4} \nonumber
&n_2^{HS}=[{\overline{\eta}_{11}^{(r)}} \, (\overline{\mu}_2^{(r)} \, \hat{I}_2
+ \overline{\eta}_{21}^{(r)} \, n_1^{HE})]^{-1} \, [(1-2 \beta)^2 \,
\overline{\eta}_{11}^2 \, (n_1^{HE})^2-(2 \beta-1) \, \overline{\eta}_{11}
\, \overline{\mu}_1^{(r)} \, \hat{I}_1 \, n_1^{HE}] \\
&-n_2^S
\end{eqnarray}
In order to have $n_1^{A} >0$, the following condition must be fulfilled:

\begin{equation} \label{condition1}
n_1^{HE} > \frac{\displaystyle \overline{\mu}_1^{(r)} \, \hat{I}_1}
{\displaystyle (2 \beta-1) \, \overline{\eta}_{11}}
\end{equation}
with $\beta > \frac{1}{2}$.
Furthermore, $n_2^S$ given by Eq. (\ref{soluzione3}) is well-defined if
$\alpha \ne 1$.

We note that if

$$n_1^{HE}=\frac{\displaystyle \overline{\mu}_1^{(r)} \, \hat{I}_1}
{\displaystyle (2 \beta-1) \, \overline{\eta}_{11}}$$
then Eqs.(\ref{soluzione2})-(\ref{soluzione4}) give
$n_1^{A}=n_2^S=n_2^{HS}=0.$
Therefore, from Eq. (\ref{2.5}) it follows $n_1^{HE}=N$.
This means that condition (\ref{condition1}) is equivalent to

$$N > \frac{\displaystyle \overline{\mu}_1^{(r)} \, \hat{I}_1}
{\displaystyle (2 \beta-1) \, \overline{\eta}_{11}}.$$

\subsection{Mean state of the populations } \label{sub_eq2}

We analyze also the equilibrium solutions for the mean epidemiological
state of each population.
From Eq. (\ref{eqU1Amedio}) it follows:

\begin{equation} \label{5.1}
\hat{U}_1^A=const.
\end{equation}
Then, setting the left-hand side of Eqs. (\ref{eqU1HEmedio}),
(\ref{eqU2Smedio}), (\ref{eqU2HSmedio}) equal to zero, and
imposing that $n_1^{HE} >0$, $n_1^{A} >0$, $n_2^{S} >0$, $n_2^{HS} >0$,
$\alpha \ne 1$, $\overline{\mu}_2^{(r)} \ne 0$, $\hat{I}_2 \ne 0$,
we get

\begin{equation} \label{5.2}
\hat{U}_2^{HS}=-\hat{U}_2^{S}
\end{equation}
\begin{equation} \label{5.3}
\hat{U}_1^{A}=-\hat{U}_2^{S}
\end{equation}
\begin{equation} \label{5.4}
\hat{U}_1^{HE}=\hat{U}_2^{S}
\end{equation}
Rearranging the results given by Eqs. (\ref{5.2})-(\ref{5.4}), one obtains:

\begin{equation} \label{5.5}
\hat{U}_1^{HE}+\hat{U}_1^A=0
\end{equation}
\begin{equation} \label{5.6}
\hat{U}_2^{S}+\hat{U}_2^{HS}=0
\end{equation}
Eqs. (\ref{5.1}), (\ref{5.5}), (\ref{5.6}) allow us to infer that,
when the equilibrium is reached,
there is a perfect balance in terms of epidemiological state for the
two groups of individuals within each population.

\section{Stability of equilibrium }
\label{stability}

In order to study the stability of the non-trivial equilibrium solutions
for the population sizes, let us consider Eqs. (\ref{dens.6})-(\ref{dens.7})
which form a closed system once we substitute $n_2$ with the following
expression:

\begin{equation} \label{stab.1}
n_2=N-n_1^{HE}-n_1^{A}
\end{equation}
where $N$ is the constant total number of individuals.
Then, we can rewrite these equations in the form:

\begin{eqnarray} \label{stab.2}
\begin{cases}
\frac{\displaystyle d n_1^{HE}}{\displaystyle d t }=f(n_1^{HE}, n_1^{A}) \\

\\
\frac{\displaystyle d n_1^{A}}{\displaystyle d t }=g(n_1^{HE}, n_1^{A})
\end{cases}
\end{eqnarray}
and apply the following

\newtheorem{theorem}{Theorem}[section]

\begin{theorem} \label{t6.1}
Let us suppose that $(\tilde{n}_1^{HE}, \tilde{n}_1^{A})$ is an equilibrium
point of the system (\ref{stab.2}), (under the hypothesis, $f,g \in C^1$)
and that the Jacobian matrix:

\begin{equation*} \label{stab.3}
J=\left(
\begin{array}{cc}
\frac{\displaystyle \partial f}{\displaystyle \partial n_1^{HE}} &
\frac{\displaystyle \partial f}{\displaystyle \partial n_1^{A}} \\
\frac{\displaystyle \partial g}{\displaystyle \partial n_1^{HE}} &
\frac{\displaystyle \partial g}{\displaystyle \partial n_1^{A}}
\end{array} \right)
\end{equation*}
evaluated at $(\tilde{n}_1^{HE}, \tilde{n}_1^{A})$, is not singular (that is,
$det(J) \neq 0$), then:

\par\noindent
(i) the equilibrium point $(\tilde{n}_1^{HE}, \tilde{n}_1^{A})$ is (locally)
asymptotically stable if all the eigenvalues of $J$ have strictly negative
real part;

\par\noindent
(ii) the equilibrium point $(\tilde{n}_1^{HE}, \tilde{n}_1^{A})$ is
unstable if at least one eigenvalue of $J$ has strictly positive real part.

\end{theorem}

\newtheorem{remark}{Remark}[section]

\begin{remark}

Theorem \ref{t6.1} holds in general for a system of $m$ equations.
Indeed, in our planar case $m=2$, the stability conditions reported in
Theorem \ref{t6.1} are equivalent to stating that an equilibrium point is
(locally) asymptotically stable if and only if: $Tr(J) <0$ and $det(J) >0$,
where $Tr(J)$ and $det(J)$ stand for 'trace' and 'determinant' of $J$,
respectively.

\end{remark}

Therefore, in order to analyze the stability of the equilibrium solution
given in Section \ref{sub_eq1}, we need to evaluate the sign of
$Tr(J)$ and $det(J)$.

Taking into account that:

\begin{eqnarray} \label{stab.4} \nonumber
\frac{\displaystyle \partial f}{\displaystyle \partial n_1^{HE}}
(\tilde{n}_1^{HE}, \tilde{n}_1^{A})&=(1-2 \beta) \, \overline{\eta}_{11} \,
\tilde{n}_1^{A}+\overline{\eta}_{21}^{(r)} \, (N-\tilde{n}_1^{HE}-
\tilde{n}_1^{A}) \\
&-\overline{\eta}_{21}^{(r)} \, \tilde{n}_1^{HE}-\overline{\mu}_2^{(r)} \,
\hat{I}_2
\end{eqnarray}

\begin{equation} \label{stab.5}
\frac{\displaystyle \partial f}{\displaystyle \partial n_1^{A}}
(\tilde{n}_1^{HE}, \tilde{n}_1^{A})=(1-2 \beta) \, \overline{\eta}_{11} \,
\tilde{n}_1^{HE}
-\overline{\eta}_{21}^{(r)} \, \tilde{n}_1^{HE}-\overline{\mu}_2^{(r)} \,
\hat{I}_2
\end{equation}

\begin{equation} \label{stab.6}
\frac{\displaystyle \partial g}{\displaystyle \partial n_1^{HE}}
(\tilde{n}_1^{HE}, \tilde{n}_1^{A})=(2 \beta-1) \, \overline{\eta}_{11} \,
\tilde{n}_1^{A}
\end{equation}

\begin{equation} \label{stab.7}
\frac{\displaystyle \partial g}{\displaystyle \partial n_1^{A}}
(\tilde{n}_1^{HE}, \tilde{n}_1^{A})=(2 \beta-1) \, \overline{\eta}_{11} \,
\tilde{n}_1^{HE}-2 \, \overline{\eta}_{11}^{(r)} \, \tilde{n}_1^{A}
-\overline{\mu}_1^{(r)} \, \hat{I}_1
\end{equation}
we get

\begin{eqnarray} \label{stab.8} \nonumber
&Tr(J)=\frac{\displaystyle \partial f}{\displaystyle \partial n_1^{HE}}
(\tilde{n}_1^{HE}, \tilde{n}_1^{A})+
\frac{\displaystyle \partial g}{\displaystyle \partial n_1^{A}}
(\tilde{n}_1^{HE}, \tilde{n}_1^{A})
=-(2 \beta-1) \frac{\displaystyle \overline{\eta}_{11}}
{\displaystyle \overline{\eta}_{11}^{(r)}}
\bigg[ (2 \beta-1) \, \overline{\eta}_{11} \, \tilde{n}_1^{HE}-
\overline{\mu}_1^{(r)} \, \hat{I}_1 \bigg] \\ \nonumber
&+\frac{\displaystyle (2 \beta-1) \, \overline{\eta}_{21}^{(r)} \,
\overline{\eta}_{11} \,
\tilde{n}_1^{HE}}{\displaystyle \overline{\eta}_{11}^{(r)}
(\overline{\eta}_{21}^{(r)} \,
\tilde{n}_1^{HE}+\overline{\mu}_2^{(r)} \, \hat{I}_2)}
\bigg[ (2 \beta-1) \, \overline{\eta}_{11} \, \tilde{n}_1^{HE}-
\overline{\mu}_1^{(r)} \, \hat{I}_1 \bigg]
-\overline{\eta}_{21}^{(r)} \, \tilde{n}_1^{HE}-\overline{\mu}_2^{(r)} \,
\hat{I}_2 \\
&-\bigg[ (2 \beta-1) \, \overline{\eta}_{11} \, \tilde{n}_1^{HE}-
\overline{\mu}_1^{(r)} \, \hat{I}_1 \bigg]
\end{eqnarray}
Since in Eq. (\ref{stab.8}) the terms in the square brackets must be
positive in equilibrium conditions and $\beta > \frac{\displaystyle 1}
{\displaystyle 2}$, then $Tr(J) <0$ if the following inequality holds:

\begin{eqnarray} \label{stab.9} \nonumber
&(2 \beta-1) \, \overline{\eta}_{11} \, \overline{\mu}_2^{(r)} \,
\hat{I}_2 \, x+(2 \beta-1) \, \overline{\eta}_{11} \,
\overline{\eta}_{21}^{(r)} \, \tilde{n}_1^{HE} \, x \\ \nonumber
&+\overline{\eta}_{11}^{(r)} \bigg( \overline{\eta}_{21}^{(r)} \,
\tilde{n}_1^{HE}+\overline{\mu}_2^{(r)} \, \hat{I}_2 \bigg) x+
\overline{\eta}_{11}^{(r)} \bigg( \overline{\eta}_{21}^{(r)} \,
\tilde{n}_1^{HE}+\overline{\mu}_2^{(r)} \, \hat{I}_2 \bigg)^2 \\
&> (2 \beta-1) \, \overline{\eta}_{21}^{(r)} \,
\overline{\eta}_{11} \, \tilde{n}_1^{HE} \, x
\end{eqnarray}
where

\begin{equation} \label{stab.10}
x=(2 \beta-1) \, \overline{\eta}_{11} \, \tilde{n}_1^{HE}-
\overline{\mu}_1^{(r)} \, \hat{I}_1 >0.
\end{equation}
By observing that all the terms on the left-hand side are positive and that
the second term is equal to the one on the right-hand side, we can
conclude that the inequality (\ref{stab.9}) is always verified.

Analogously, we can compute:

\begin{eqnarray} \label{stab.11} \nonumber
&det(J)=\frac{\displaystyle \partial f}{\displaystyle \partial n_1^{HE}}
(\tilde{n}_1^{HE}, \tilde{n}_1^{A}) \,
\frac{\displaystyle \partial g}{\displaystyle \partial n_1^{A}}
(\tilde{n}_1^{HE}, \tilde{n}_1^{A})-
\frac{\displaystyle \partial f}{\displaystyle \partial n_1^{A}}
(\tilde{n}_1^{HE}, \tilde{n}_1^{A}) \,
\frac{\displaystyle \partial g}{\displaystyle \partial n_1^{HE}}
(\tilde{n}_1^{HE}, \tilde{n}_1^{A}) \\ \nonumber
&=\bigg \{-(2 \beta-1) \, \overline{\eta}_{11} \,\tilde{n}_1^{A}+
\overline{\eta}_{21}^{(r)} \, (N-\tilde{n}_1^{HE}-\tilde{n}_1^{A})-
[\overline{\eta}_{21}^{(r)} \, \tilde{n}_1^{HE}+\overline{\mu}_2^{(r)} \,
\hat{I}_2] \bigg \}  \\ \nonumber
& \cdot \bigg \{ [ (2 \beta-1) \, \overline{\eta}_{11} \, \tilde{n}_1^{HE}-
\overline{\mu}_1^{(r)} \, \hat{I}_1]-2 \, \overline{\eta}_{11}^{(r)}\,
\tilde{n}_1^{A} \bigg \} \\ \nonumber
&+\bigg \{ (2 \beta-1) \, \overline{\eta}_{11} \, \tilde{n}_1^{HE}+
[\overline{\eta}_{21}^{(r)} \, \tilde{n}_1^{HE}+
\overline{\mu}_2^{(r)} \, \hat{I}_2] \bigg \} \, (2 \beta-1) \,
\overline{\eta}_{11} \, \tilde{n}_1^{A} \\
&=T_1+T_2
\end{eqnarray}
where

\begin{eqnarray} \label{stab.12} \nonumber
&T_1=\bigg \{-(2 \beta-1) \, \overline{\eta}_{11} \,\tilde{n}_1^{A}+
\overline{\eta}_{21}^{(r)} \, (N-\tilde{n}_1^{HE}-\tilde{n}_1^{A})-
[\overline{\eta}_{21}^{(r)} \, \tilde{n}_1^{HE}+\overline{\mu}_2^{(r)} \,
\hat{I}_2] \bigg \}  \\
& \cdot \bigg \{ [ (2 \beta-1) \, \overline{\eta}_{11} \, \tilde{n}_1^{HE}-
\overline{\mu}_1^{(r)} \, \hat{I}_1]-2 \, \overline{\eta}_{11}^{(r)}\,
\tilde{n}_1^{A} \bigg \}
\end{eqnarray}
\begin{equation} \label{stab.13}
T_2=\bigg \{ (2 \beta-1) \, \overline{\eta}_{11} \, \tilde{n}_1^{HE}+
[\overline{\eta}_{21}^{(r)} \, \tilde{n}_1^{HE}+
\overline{\mu}_2^{(r)} \, \hat{I}_2] \bigg \} \, (2 \beta-1) \,
\overline{\eta}_{11} \, \tilde{n}_1^{A}
\end{equation}
Since $T_2 >0$, in order to prove that $det(J) >0$, we have to only
evaluate the sign of $T_1$:

\begin{equation} \label{stab.14}
T_1=(2 \beta-1) \,\frac{\displaystyle \overline{\eta}_{11}}
{\displaystyle \overline{\eta}_{11}^{(r)}} \, x^2-
\frac{\displaystyle (2 \beta-1) \, \overline{\eta}_{21}^{(r)} \,
\overline{\eta}_{11} \,
\tilde{n}_1^{HE}}{\displaystyle \overline{\eta}_{11}^{(r)}
(\overline{\eta}_{21}^{(r)} \,
\tilde{n}_1^{HE}+\overline{\mu}_2^{(r)} \, \hat{I}_2)} x^2
+(\overline{\eta}_{21}^{(r)} \, \tilde{n}_1^{HE}+
\overline{\mu}_2^{(r)} \, \hat{I}_2) \, x
\end{equation}
where $x$ is a positive quantity given by (\ref{stab.10}).
From (\ref{stab.14}) we deduce that $T_1 >0$ if

\begin{eqnarray} \label{stab.15} \nonumber
&(2 \beta-1) \, \overline{\eta}_{11} \, \overline{\mu}_2^{(r)} \, \hat{I}_2
\, x+(2 \beta-1) \, \overline{\eta}_{11} \, \overline{\eta}_{21}^{(r)} \,
\tilde{n}_1^{HE} \, x
+\overline{\eta}_{11}^{(r)} \, \bigg( \overline{\eta}_{21}^{(r)} \,
\tilde{n}_1^{HE}+\overline{\mu}_2^{(r)} \, \hat{I}_2 \bigg)^2 \\
&> (2 \beta-1) \, \overline{\eta}_{11} \,
\overline{\eta}_{21}^{(r)} \,\tilde{n}_1^{HE} \, x
\end{eqnarray}
Since all the terms on the left-hand side are positive and the second
term is equal to the one on the right-hand side, we infer that the
inequality (\ref{stab.15}) is always verified.

On the other hand, evaluating the expressions (\ref{stab.4})-(\ref{stab.7})
in correspondence of the trivial equilibrium solution
$(n_1^{HE},n_1^{A})=(N,0)$, we easily get that both eigenvalues of the
Jacobian matrix are negative (the equilibrium is stable) if and only if

$$N < \frac{\displaystyle \overline{\mu}_1^{(r)} \, \hat{I}_1}
{\displaystyle (2 \beta-1) \, \overline{\eta}_{11}}$$
that is, when the endemic equilibrium does not exist.

\section{Numerical simulations} \label{numsim}

In the following we present some numerical test cases in order to assess the
ability of our model to qualitatively reproduce the main features of
SARS-CoV-2 spread.
We have numerically integrated Eqs. (\ref{dens.6}) and (\ref{dens.7}) until
an equilibrium solution has been obtained.
By taking into account that the total number of individuals $N$ is conserved,
we have substituted

\begin{equation} \label{7.0}
n_2=N-n_1^{HE}-n_1^A
\end{equation}
in Eq. (\ref{dens.6}).
Thus, Eqs. (\ref{dens.6}) and (\ref{dens.7}) form a closed system of equations
that can be solved to give the temporal evolution of the number densities of
healthy individuals ($n_1^{HE}$) and of asymptomatic carriers ($n_1^A$).
In all the numerical simulations presented below we have fixed $N=1$.
To estimate the numerical error, we have compared the equilibrium solution
obtained by integrating Eqs. (\ref{dens.6}) and (\ref{dens.7}) and the one
computed analytically using the formulas reported in Section \ref{sub_eq1}
(where the terms have been rearranged in order to insert in
Eq. (\ref{soluzione4}) the expression (\ref{7.0}) with $N=1$).
The agreement has proved to be very good in all cases considered with the
error always within $0.1\%$.

\bigskip

\noindent
{\bf Test case (1).}

Here we investigate the impact of an extremely contagious variant by setting
$\beta=1$ in the stochastic law modeling the interactions between a healthy
individual and an asymptomatic carrier (formula (\ref{lab})). \\

\begin{figure}[!httt]
\centering
\subfigure{(a)\includegraphics[trim={0.0cm 0.21cm 0.0cm 0.5cm}, clip, width=0.64\textwidth, valign=t]{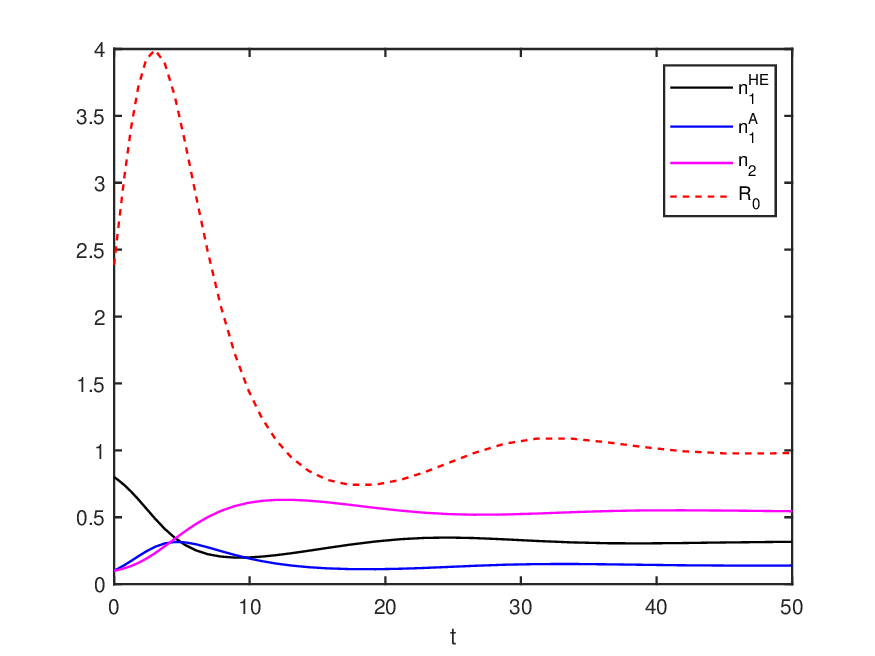}}
\subfigure{(b)\includegraphics[trim={0.0cm 0.21cm 0.0cm 0.5cm}, clip, width=0.64\textwidth, valign=t]{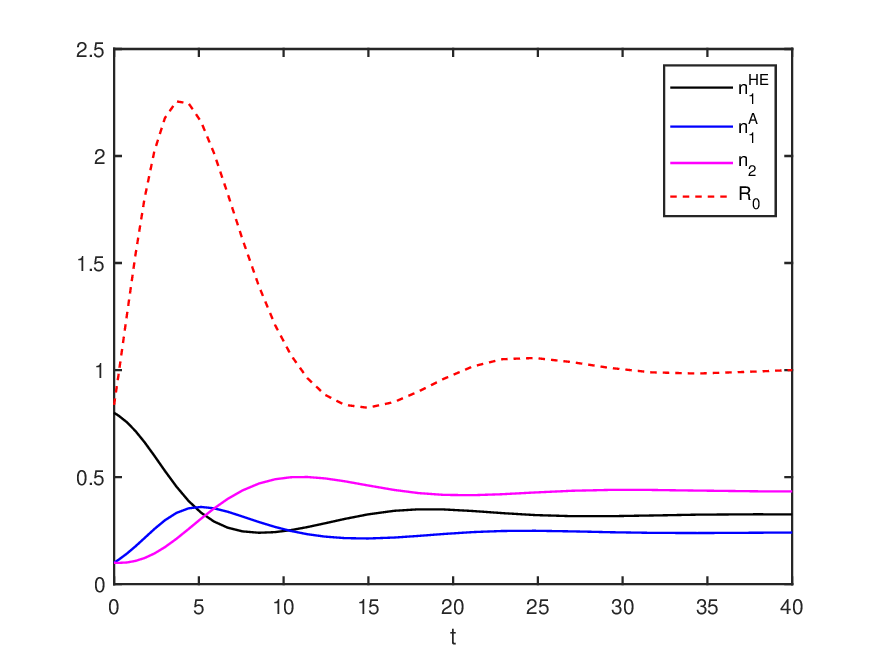}}
\subfigure{(c)\includegraphics[trim={0.0cm 0.23cm 0.0cm 0.5cm}, clip, width=0.64\textwidth, valign=t]{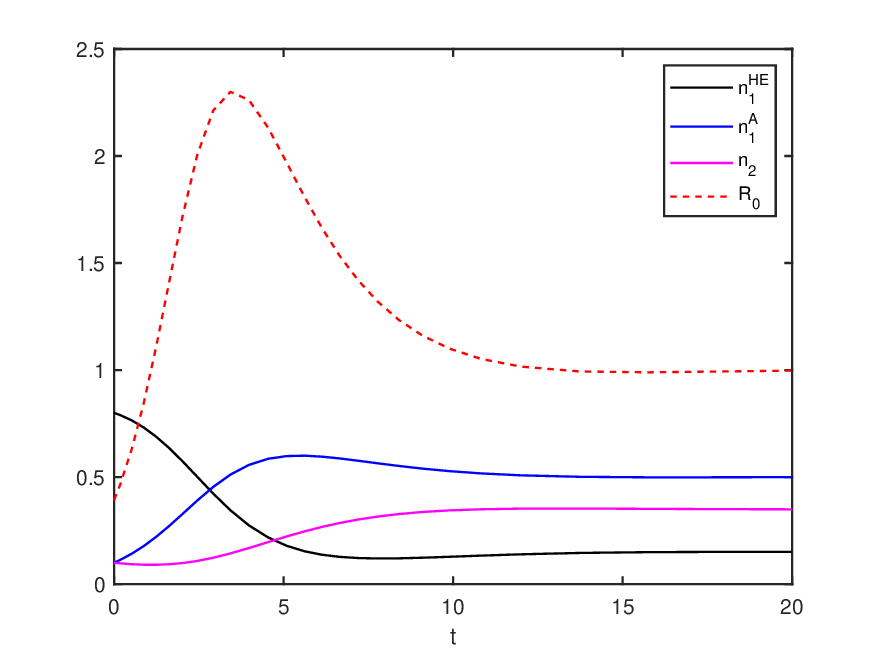}}
\caption{{\footnotesize Time evolution of the initial data $(n_1^{HE}, n_1^A, n_2)(t=0) = (0.8, 0.1, 0.1)$ and of the reproduction number $R_0$ with $\beta=1$, $\bar{\eta}_{11}=1$, $\bar{\eta}_{11}^{(r)}=0.1$.
Panel (a): $\bar{\eta}_{21}^{(r)}=0.1$, $\bar{\mu}_1^{(r)}\, \hat{I}_1 = 0.3$, $\bar{\mu}_2^{(r)}\, \hat{I}_2 = 0.05$.
Panel (b): $\bar{\eta}_{21}^{(r)}=0.4$, $\bar{\mu}_1^{(r)}\, \hat{I}_1 = 0.3$, $\bar{\mu}_2^{(r)}\, \hat{I}_2 = 0.05$.
Panel (c): $\bar{\eta}_{21}^{(r)}=0.1$, $\bar{\mu}_1^{(r)}\, \hat{I}_1 = 0.1$, $\bar{\mu}_2^{(r)}\, \hat{I}_2 = 0.2$.
}}
\label{figure1}
\end{figure}

\begin{figure}[!httt]
\centering
\subfigure{(a)\includegraphics[trim={0.0cm 0.3cm 0.0cm 1.4cm}, clip, width=0.69\textwidth, valign=t]{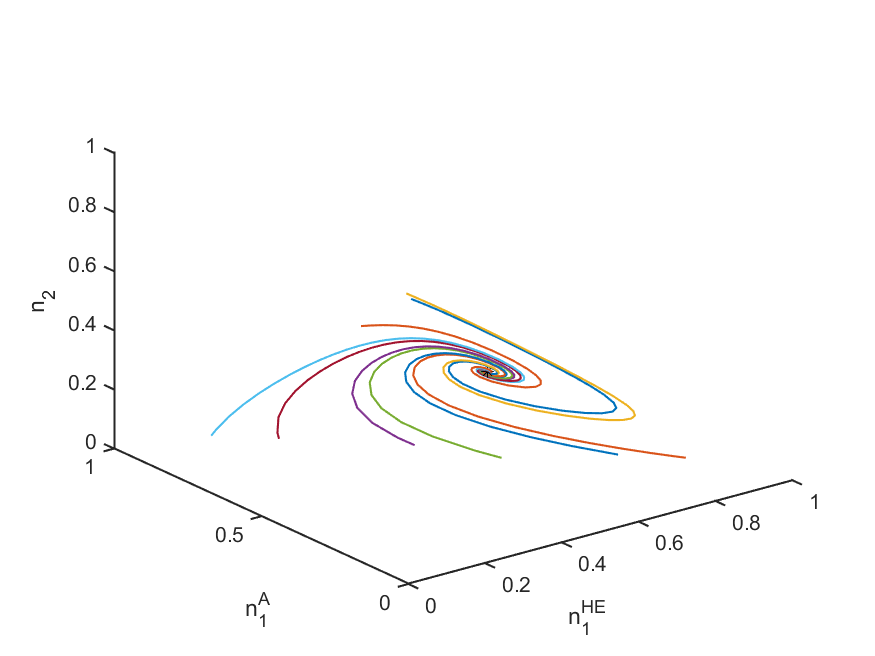}}
\subfigure{(b)\includegraphics[trim={0.0cm 0.3cm 0.0cm 1.6cm}, clip, width=0.69\textwidth, valign=t]{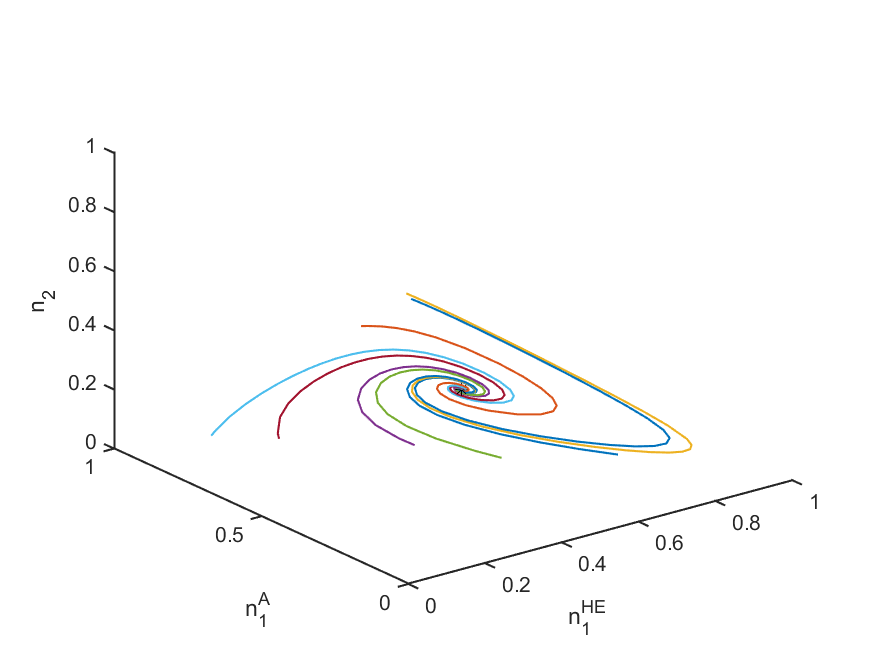}}
\subfigure{(c)\includegraphics[trim={0.0cm 0.3cm 0.0cm 1.6cm}, clip, width=0.69\textwidth, valign=t]{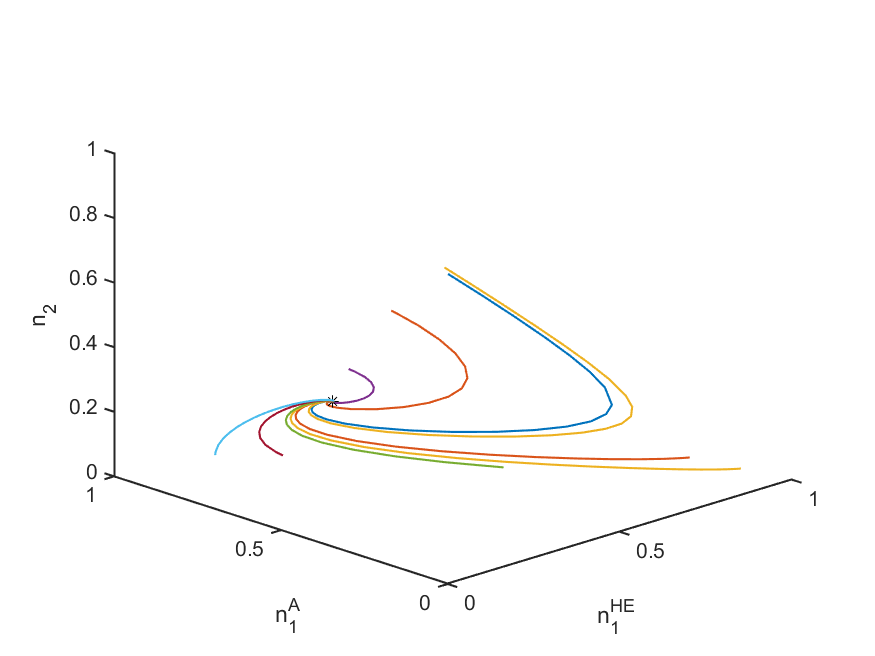}}
\caption{{\footnotesize Phase portrait of the evolution of $n_1^{HE}$, $n_1^A$, $n_2$ for a variant with contagion rate $\beta=1$, and $\bar{\eta}_{11}=1$, $\bar{\eta}_{11}^{(r)}=0.1$.
Panel (a): $\bar{\eta}_{21}^{(r)}=0.1$, $\bar{\mu}_1^{(r)}\, \hat{I}_1 = 0.3$, $\bar{\mu}_2^{(r)}\, \hat{I}_2 = 0.05$.
Panel (b): $\bar{\eta}_{21}^{(r)}=0.4$, $\bar{\mu}_1^{(r)}\, \hat{I}_1 = 0.3$, $\bar{\mu}_2^{(r)}\, \hat{I}_2 = 0.05$.
Panel (c): $\bar{\eta}_{21}^{(r)}=0.1$, $\bar{\mu}_1^{(r)}\, \hat{I}_1 = 0.1$, $\bar{\mu}_2^{(r)}\, \hat{I}_2 = 0.2$.
}}
\label{figure2}
\end{figure}

{\it (1.a)}
First, we consider a situation where social restrictions are not applied.
Therefore, we assume that interactions between individuals without symptoms
(described by the parameter $\bar{\eta}_{11}$) are much more probable than
others:

$$
\bar{\eta}_{11}=1\,, \qquad \quad \bar{\eta}_{11}^{(r)}=0.1\,,
\qquad \quad \bar{\eta}_{21}^{(r)}=0.1.
$$
Furthermore, we take into account the action of the immune system by
choosing:

$$
\bar{\mu}_1^{(r)}\, \hat{I}_1 = 0.3\,, \qquad \quad
\bar{\mu}_2^{(r)}\, \hat{I}_2 = 0.05\,.
$$
Starting from the initial conditions

\begin{equation} \label{7.1}
(n_1^{HE}, n_1^A, n_2)(t=0) = (0.8, 0.1, 0.1)
\end{equation}
Eqs. (\ref{dens.6}) and (\ref{dens.7}) have been integrated and the time
evolution of the number density of healthy individuals ($n_1^{HE}$),
positive asymptomatic ($n_1^A$) and ill people ($n_2$) is reported in
Fig. \ref{figure1}-(a).
In the same picture the profile of the time-dependent effective reproduction
number $R_0$ (Eq. (\ref{r0})) is also shown in order to quantify the
effectiveness of virus containment strategies.
Since for the chosen parameters the condition (\ref{condition1}) is fulfilled,
our
system of equations (\ref{dens.6}) and (\ref{dens.7}) admits a non-trivial
equilibrium solution given by

\begin{equation} \label{7.2}
(n_1^{HE}, n_1^A, n_2) = (0.3141, 0.1411, 0.5448).
\end{equation}
In the framework of an epidemiological analysis, it is of paramount importance
to also evaluate how an endemic equilibrium is achieved.
Therefore, in Figure \ref{figure2}-(a) the phase portrait, corresponding to the
system (\ref{dens.6})-(\ref{dens.7}), is shown in the three-dimensional
space ($n_1^{HE}, n_1^A, n_2$).
From this picture, a characteristic spiral shape of the curves is evident
with peaks and troughs of infections.
This behavior closely resembles the concept of an epidemic wave.
Indeed, the SARS-CoV-2 pandemic developed as a series of waves: surges of
new infections followed by declines.
Although there is no a common idea among researchers of what constitutes
an epidemic wave, a useful working definition has been proposed in
\cite{ZMGW},
where this concept has been linked to the profile of the effective
reproduction number $R_0$.
If $R_0$ is larger than $1$ for a sustained period, one can identify this
time as an upward period for the epidemic wave.
On the contrary, if $R_0$ is smaller than $1$ for a sustained period, this
time can be identified as a downward period for the epidemic wave.
Relying on this definition, one can infer from Fig. \ref{figure1}-(a) the existence
of two pandemic waves as suggested also by the phase portrait in Fig. \ref{figure2}-(a).

\medskip

\noindent
{\it (1.b)}
In the following, with respect to the test case {\it (1.a)}, we consider an
epidemic scenario in which treatments against the disease are more effective.
Therefore, leaving all other parameters unchanged, we increase the value
of $\bar{\eta}_{21}^{(r)}$ (which describes the rate of interaction between
the medical staff and the infected individuals with symptoms) by setting
$\bar{\eta}_{21}^{(r)}=0.4$.
Indeed, in the context of our model, a large value of this parameter indicates
a high transition rate from population $2$ to population $1$ (healthy
individuals).
Again, starting from the initial conditions

$$
(n_1^{HE}, n_1^A, n_2)(t=0) = (0.8, 0.1, 0.1),
$$
Eqs. (\ref{dens.6}) and (\ref{dens.7}) have been integrated and the time
evolution of the number densities $n_1^{HE}, n_1^A, n_2$ is plotted in
Fig. \ref{figure1}-(b) along with the profile of the time-dependent effective
reproduction number $R_0$.
For this test case, the non-trivial equilibrium solution is given by

\begin{equation} \label{7.3}
(n_1^{HE}, n_1^A, n_2) = (0.3241, 0.2412, 0.4347).
\end{equation}
As expected, the number of sick people is lower than in the previous case
{\it (1.a)} (see expression (\ref{7.2})).
To complete the scenario, the phase portrait is shown in Figure \ref{figure2}-(b).
From these pictures, we note that the duration of the pandemic is shorter
and that the peak of $R_0$ is lower than in the test case {\it (1.a)},
while the phase portraits are quite similar.

\medskip

\noindent
{\it (1.c)}
In this test, we assess the impact of vaccination, which affects the response
of both the innate and adaptive immune system.
In particular, with respect to the test case {\it (1.a)}, we increase the
term ($\bar{\mu}_2^{(r)}\, \hat{I}_2$) (which describes the overall response
of the adaptive immune system on individuals belonging to population $2$)
and we decrease the term ($\bar{\mu}_1^{(r)}\, \hat{I}_1$) (which models
the average action of the innate immune system on asymptomatic individuals)
by setting

$$
\bar{\mu}_1^{(r)}\, \hat{I}_1 = 0.1\,, \qquad \quad
\bar{\mu}_2^{(r)}\, \hat{I}_2 = 0.2\,.
$$
In the framework of the model we derived, large values of the term
($\bar{\mu}_2^{(r)}\, \hat{I}_2$) indicate that a significant fraction of sick
people recovers without specific treatments, while small values of the
term ($\bar{\mu}_1^{(r)}\, \hat{I}_1$) denote a negligible probability for
the hyperinflammation phenomenon to occur.
Starting from the same initial conditions previously considered in the
tests {\it (1.a)} and {\it (1.b)},

$$
(n_1^{HE}, n_1^A, n_2)(t=0) = (0.8, 0.1, 0.1),
$$
Eqs. (\ref{dens.6}) and (\ref{dens.7}) have been integrated and the time
evolution of the number densities $n_1^{HE}, n_1^A, n_2$ along with the
trend of the effective reproduction number $R_0$ are shown in Fig. \ref{figure1}-(c).
For the set of parameters considered, our system of equations admits a
non-trivial equilibrium solution given by

\begin{equation} \label{7.4}
(n_1^{HE}, n_1^A, n_2) = (0.1501, 0.5010, 0.3489).
\end{equation}
Compared to previous tests, the number of ill persons belonging to
population $2$ is significantly reduced.
The large fraction of carriers can be explained by taking into account that
we are analyzing an extremely contagious variant ($\beta=1$) in which each
healthy individual interacting with a positive asymptomatic becomes infectious.
Therefore, due to the large number of healthy people, also the number of
positive asymptomatic persons will be higher than in previous tests.
Furthermore, this trend is perfectly in line with the realistic situation
created by the administration of Covid-19 vaccines, which have reduced the
number of ill individuals without however preventing people from contracting
the virus.
Figure \ref{figure1}-(c) shows that the convergence to equilibrium is faster than in
previous tests without subsequent epidemic waves, as confirmed by the phase
portrait given in Figure \ref{figure2}-(c).

\medskip

\noindent
{\bf Test case (2).}

In the following we analyze the evolution of an epidemic driven by a variant
with a lower transmission rate than in the {\bf test case (1)}.
Therefore, we set $\beta=0.7$.\\

\begin{figure}[!httt]
\centering
\subfigure{(a)\includegraphics[trim={0.0cm 0.21cm 0.0cm 0.5cm}, clip, width=0.64\textwidth, valign=t]{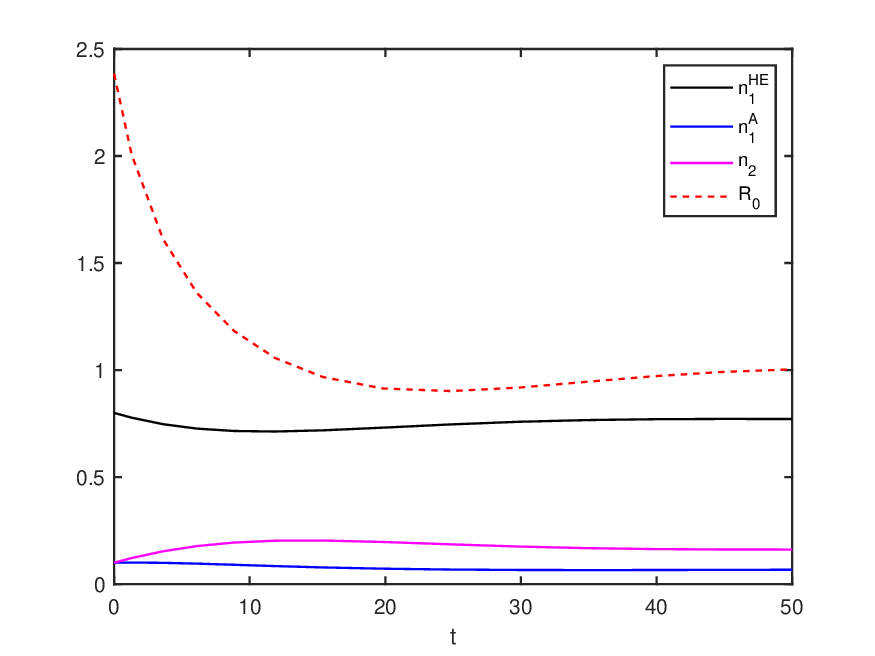}}
\subfigure{(b)\includegraphics[trim={0.0cm 0.21cm 0.0cm 0.5cm}, clip, width=0.64\textwidth, valign=t]{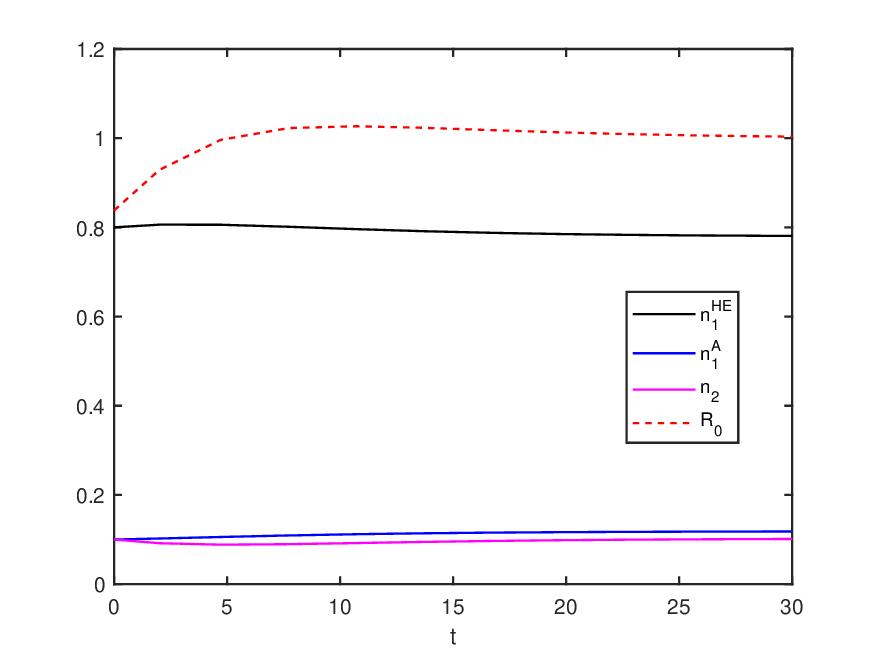}}
\subfigure{(c)\includegraphics[trim={0.0cm 0.23cm 0.0cm 0.5cm}, clip, width=0.64\textwidth, valign=t]{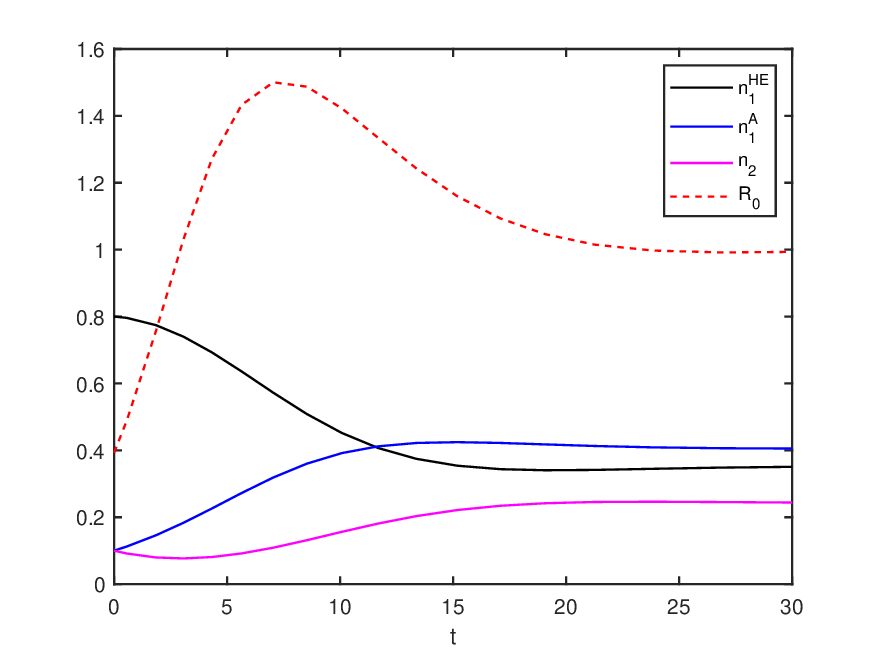}}
\caption{{\footnotesize Time evolution of the initial data $(n_1^{HE}, n_1^A, n_2)(t=0) = (0.8, 0.1, 0.1)$ and of the reproduction number $R_0$ with $\beta=0.7$, $\bar{\eta}_{11}=1$, $\bar{\eta}_{11}^{(r)}=0.1$.
Panel (a): $\bar{\eta}_{21}^{(r)}=0.1$, $\bar{\mu}_1^{(r)}\, \hat{I}_1 = 0.3$, $\bar{\mu}_2^{(r)}\, \hat{I}_2 = 0.05$.
Panel (b): $\bar{\eta}_{21}^{(r)}=0.4$, $\bar{\mu}_1^{(r)}\, \hat{I}_1 = 0.3$, $\bar{\mu}_2^{(r)}\, \hat{I}_2 = 0.05$.
Panel (c): $\bar{\eta}_{21}^{(r)}=0.1$, $\bar{\mu}_1^{(r)}\, \hat{I}_1 = 0.1$, $\bar{\mu}_2^{(r)}\, \hat{I}_2 = 0.2$.
}}
\label{figure3}
\end{figure}

\begin{figure}[!httt]
\centering
\subfigure{(a)\includegraphics[trim={0.0cm 0.1cm 0.0cm 1.2cm}, clip, width=0.68\textwidth, valign=t]{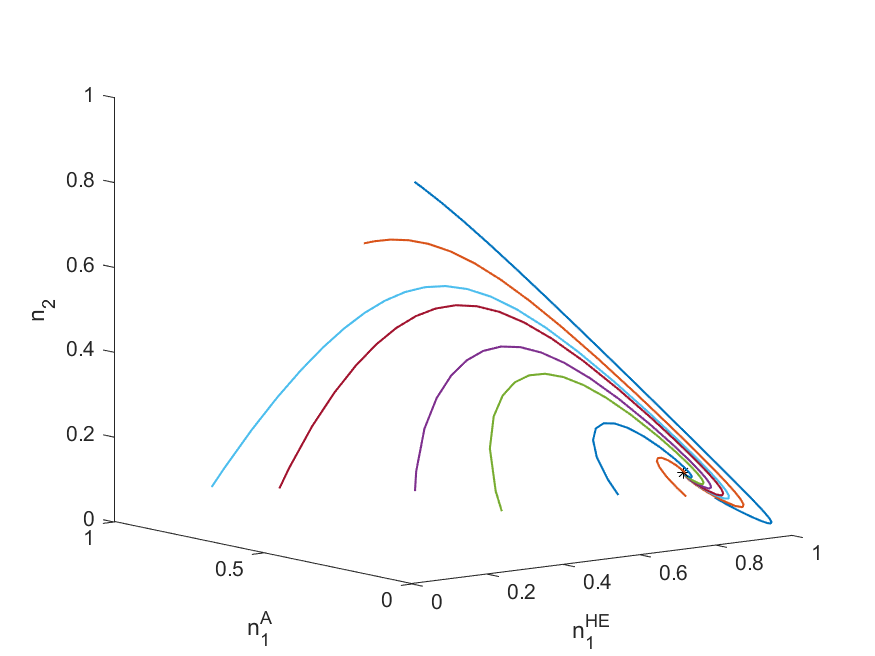}}
\subfigure{(b)\includegraphics[trim={0.0cm 0.3cm 0.0cm 1.6cm}, clip, width=0.68\textwidth, valign=t]{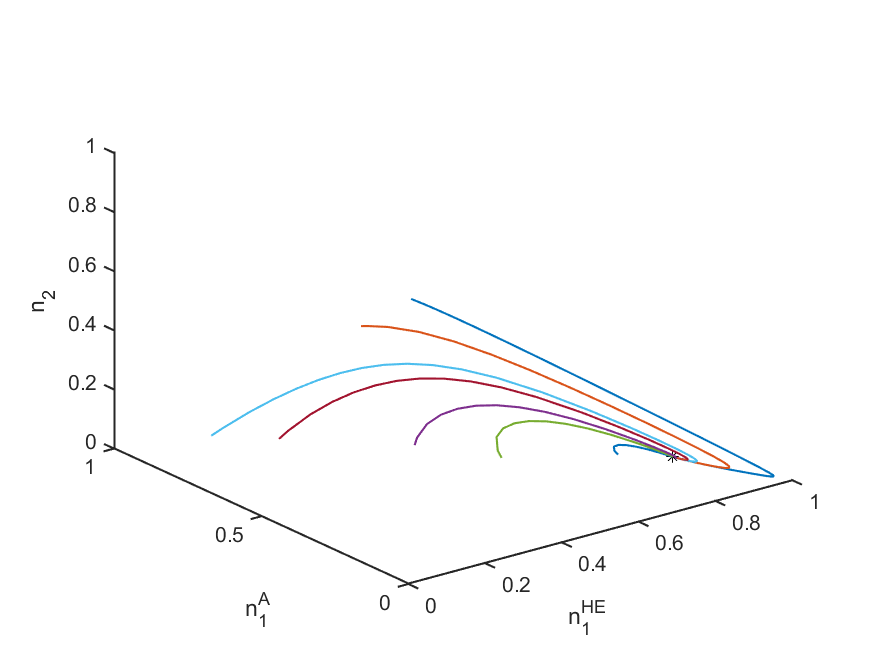}}
\subfigure{(c)\includegraphics[trim={0.0cm 0.3cm 0.0cm 1.6cm}, clip, width=0.68\textwidth, valign=t]{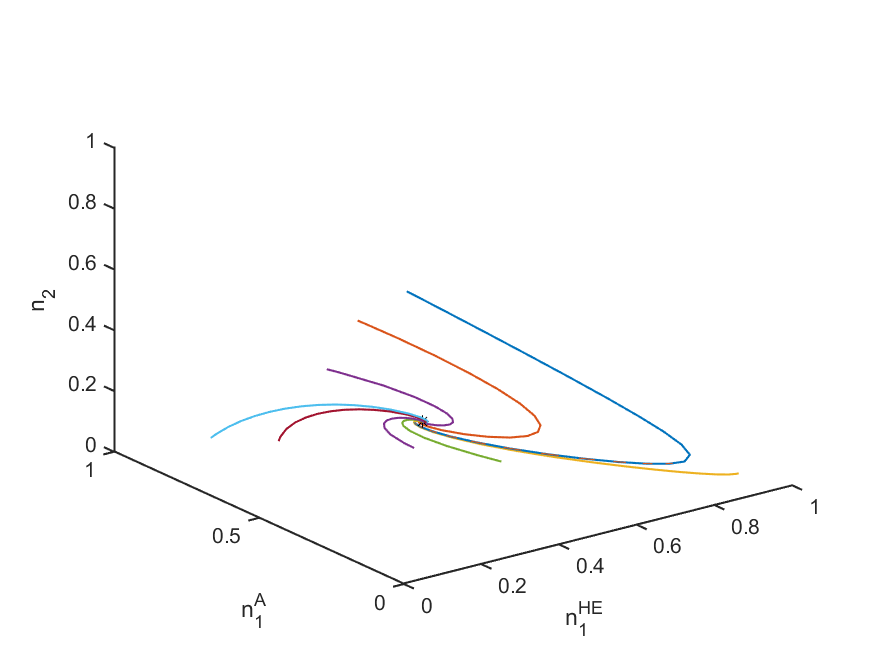}}
\caption{{\footnotesize Phase portrait of the evolution of $n_1^{HE}$, $n_1^A$, $n_2$ for a variant with contagion rate $\beta=0.7$, and $\bar{\eta}_{11}=1$, $\bar{\eta}_{11}^{(r)}=0.1$.
Panel (a): $\bar{\eta}_{21}^{(r)}=0.1$, $\bar{\mu}_1^{(r)}\, \hat{I}_1 = 0.3$, $\bar{\mu}_2^{(r)}\, \hat{I}_2 = 0.05$.
Panel (b): $\bar{\eta}_{21}^{(r)}=0.4$, $\bar{\mu}_1^{(r)}\, \hat{I}_1 = 0.3$, $\bar{\mu}_2^{(r)}\, \hat{I}_2 = 0.05$.
Panel (c): $\bar{\eta}_{21}^{(r)}=0.1$, $\bar{\mu}_1^{(r)}\, \hat{I}_1 = 0.1$, $\bar{\mu}_2^{(r)}\, \hat{I}_2 = 0.2$.
}}
\label{figure4}
\end{figure}

{\it (2.a)}
Here the parameters we use to integrate Eqs. (\ref{dens.6}) and (\ref{dens.7})
along with the initial conditions are the same as in test {\it (1.a)}.
In this case, the non-trivial equilibrium solution is given by

\begin{equation} \label{7.5}
(n_1^{HE}, n_1^A, n_2) = (0.7671, 0.0684, 0.1646).
\end{equation}
As expected, the number of healthy individuals is much higher than in test
{\it (1.a)}, and consequently the number of sick people is much lower.
Looking at the time evolution of the number densities $n_1^{HE}, n_1^A, n_2$
and of the effective reproduction number $R_0$ shown in Fig. \ref{figure3}-(a), one
infers that there are no epidemic waves and the outbreak dies out very
quickly.
These features are also confirmed by the phase portrait reported in
Fig. \ref{figure4}-(a).

\medskip

\noindent
{\it (2.b)}
For this simulation we have chosen the same values of the parameters
and of the initial conditions as in test {\it (1.b)} in order to integrate
Eqs. (\ref{dens.6}) and (\ref{dens.7}).
The non-trivial equilibrium solution obtained is

\begin{equation} \label{7.6}
(n_1^{HE}, n_1^A, n_2) = (0.7795, 0.1182, 0.1023).
\end{equation}
From the comparison with the previous case {\it (2.a)}, it can be deduced that,
as regards variants with low transmission rates, the effectiveness of medical
treatments is not a decisive factor in reducing the number of sick people,
since the equilibrium values differ only slightly.
Indeed, the time evolution of the number densities $n_1^{HE}, n_1^A, n_2$
and of the effective reproduction number $R_0$, shown in Fig. \ref{figure3}-(b),
along with the phase portrait, presented in Fig. \ref{figure4}-(b), indicate that
the duration of the epidemic is shorter than in the case {\it (2.a)}.

\medskip

\noindent
{\it (2.c)}
In order to evaluate the effects of vaccination, we choose the same
parameters considered in test {\it (1.c)}.
Starting from the same initial conditions, the integration of
Eqs. (\ref{dens.6}) and (\ref{dens.7}) leads to the following endemic
equilibrium solution

\begin{equation} \label{7.7}
(n_1^{HE}, n_1^A, n_2) = (0.3516, 0.4063, 0.2422).
\end{equation}
The comparison of result (\ref{7.7}) with formulas (\ref{7.5}) and
(\ref{7.6}) allows one to conclude that, for variants with low transmission
rates, vaccination is not so effective, in terms of reducing the number
of sick people, as it has been observed in the case of highly contagious
variants (see test {\it (1.c)}).
However, the time evolution of the number densities $n_1^{HE}, n_1^A, n_2$
and of the effective reproduction number $R_0$, shown in Fig. \ref{figure3}-(c), along
with the phase portrait, reported in Fig. \ref{figure4}-(c), confirm that the epidemic
ends very quickly, as expected.

\medskip
Beyond the simulations presented above, we have further checked numerically
that, if we introduce lockdown measures decreasing the value of the
parameter $\bar{\eta}_{11}$  in the test cases {\it (1.a)} and {\it (2.a)},
that is the rate of interaction between individuals without symptoms, then
the number of sick persons is drastically reduced.
As a final remark, it is interesting to note how the peaks in the
temporal evolution of the effective reproduction number $R_0$ are always
in correspondence (even if not exactly coinciding) with the peaks in the
number density profile of positive asymptomatic individuals ($n_1^A$), while
the maximum value of the number density of sick people ($n_2$) is slightly
postponed in time.
This trend reflects what has been observed in the Covid-19 pandemic, where
the peak of the infection has always been followed (during the different
waves), with a slight time delay, by the peak in the number of ill persons.

%\begin{figure}[!httt]
%\centering
%\subfigure{(d)\includegraphics[trim={0.0cm 0.73cm 0.0cm 0.5cm}, clip, width=0.7\textwidth, valign=t]{R0-ni-2-A4-def-mod.eps}}
%\caption{}
%\end{figure}

\section{Conclusion} \label{final}

In this work, we have exploited the Statistical Mechanics methods in order to
model the human-to-human mechanisms of SARS-CoV-2 transmission.
Starting from a description of the interactions between individuals based on
the Boltzmann equation (microscopic scale), we have derived a set of evolution
equations for the size and mean state of each population considered, that is
healthy people, positive asymptomatic, positive symptomatic and hospitalized
persons (macroscopic scale).
Unlike the epidemiological models that have been developed within the
kinetic theory framework in the last few years, our approach does not rely
on one of the existing compartmental models, but we derive a new system of
macroscopic evolution equations, which aims to account for the characteristics
of Covid-19.
In particular, our model focuses on the role of positive-asymptomatic
individuals in triggering the spread of SARS-CoV-2 virus.
Asymptomatic transmission has been considered as the "Achilles' heel" of
Covid-19 control strategies.
Indeed, several studies have demonstrated that asymptomatic carriers have
infected a similar number of people as symptomatic individuals
\cite{BA}, \cite{Wetal}.
Beyond a trivial equilibrium solution corresponding to
$n_1^A=n_2^S=n_2^{HS}=0$ (that is, the eradication of the disease),
our model predicts, under suitable conditions on infection parameters,
the existence of an endemic
equilibrium which is always asymptotically stable.

Changing the value of some relevant parameters in the derived set of
macroscopic equations, the following aspects of the SARS-CoV-2 spread
can be addressed:

\par\noindent
(i) The impact of the contagiousness of new variants.
By setting $\beta=1$, in the stochastic law modeling the interactions between
a healthy individual and a positive asymptomatic, an extremely contagious
disease is described, while lower values of $\beta$ account for less
infectious scenarios.

\par\noindent
(ii) The effects of government restrictions aimed at limiting the mobility
of individuals.
In particular, assuming a high rate of interaction between individuals
without symptoms (e.g. $\overline{\eta}_{11}=1$) means that such limitations
are not applied, while lower values of $\overline{\eta}_{11}$ describe
lockdown situations.

\par\noindent
(iii) The effectiveness of specific medical treatments.
The key parameter in this respect is $\overline{\eta}_{21}^{(r)}$, which
describes the rate of interaction between the medical staff and the infected
people belonging to population $2$.
If one takes increasing values of $\overline{\eta}_{21}^{(r)}$, our model
equations allow to study the efficacy of new drugs in order to control
the spread of the disease until its complete (possibly) eradication.

\par\noindent
(iv) The role of the innate and adaptive immunity in setting the severity
of Covid-19.
The peculiarity of SARS-CoV-2 infection is traced back to a complex interplay
between the virus and the immune system.
It involves pathogenic cell activation leading to hyperinflammation with
a major complication of the disease.
This aspect can be taken into account in the macroscopic equations we derived
by choosing large values of the term ($\overline{\mu}_1^{(r)} \, \hat{I}_1$),
which models the average action of the innate immune system on asymptomatic
individuals (who have already contracted the virus).
Furthermore, relying on clinical data, our model provides for the possibility
that ill persons recover without specific treatments, by assuming large
values of the term ($\overline{\mu}_2^{(r)} \, \hat{I}_2$), which describes
the overall response of the adaptive immune system on individuals
belonging to population $2$.
Since adaptive immunity can be acquired through the administration of vaccines,
which in turn increase the anti-viral activity of some innate immune cells,
our system of equations allows one also to assess the impact of vaccination
by increasing the term ($\overline{\mu}_2^{(r)} \, \hat{I}_2$) and,
at the same time, decreasing ($\overline{\mu}_1^{(r)} \, \hat{I}_1$).

The numerical test cases presented in Section \ref{numsim} highlight the
ability of our model to reproduce qualitatively the salient features of
Covid-19, as experienced in very recent years.
Future research should be aimed at a quantitative comparison of the
mathematical results with the medical data collected during the recent
pandemic, in order to organize also these data in a more systematic way.

\appendix

\section*{Appendix A: Equivalent formulations of the Boltzmann collision
operator}
\label{appA}
\renewcommand{\theequation}{A.\arabic{equation}}
\setcounter{equation}{0}

In the kinetic theory of gases, based on an evolution equation of the
distribution function $f(t, \mathbf{x}, \mathbf{v})$ (depending on time
$t \in \mathbb{R}_+$, space $\mathbf{x} \in \mathbb{R}^3$, molecular velocity
$\mathbf{v} \in \mathbb{R}^3$), different collision--like operators have been
proposed in the literature, in order to describe the effects due to binary
interactions between gas particles.
Specifically, for a single monatomic gas, one can prove the equivalence of the
following formulations of the Boltzmann collision operator \cite{BPS}.
Analogous results hold also for mixtures of different monatomic constituents
\cite{SNB}.

(a) Kinetic formulation:

\begin{equation} \label{A.1}
Q(f,f)(\mathbf{v})=\int_{\mathbb{R}^3} d\mathbf{w} \int_{\mathbb{S}^2} g \,
I(g, \chi) \,
[f(\mathbf{v'}) f(\mathbf{w'})-f(\mathbf{v}) f(\mathbf{w}) ] \,
d\hat{\Omega}'.
\end{equation}
In the classical form, the Boltzmann collision term is a quadratic operator
provided by the difference between a gain and a loss contribution
\cite{Cerci}. In the gain term, distributions are depending on pre-collision
velocities $\mathbf{v'}, \mathbf{w'}$ corresponding to the post-collision
ones $\mathbf{v}, \mathbf{w}$. The relative velocity of the interacting
particles is denoted by
$$
\mathbf{g} = \mathbf{v}-\mathbf{w} = g\, \hat{\Omega}
$$
where
$$
g=\vert \mathbf{v}-\mathbf{w} \vert, \qquad \quad \hat{\Omega}=
\frac{(\mathbf{v}-\mathbf{w})}{g}\,.
$$
In the kernel of the integral operator (\ref{A.1}) there appears the so-called
differential scattering cross section $I(g, \chi)$, which depends on the
relative
speed and on the deflection angle of the relative motion
$$\chi=\arccos(\hat{\Omega} \cdot \hat{\Omega}'),$$
In addition, the explicit form of the cross section is determined by
the intermolecular interaction potential of the
considered gaseous medium.
For instance, inverse-power intermolecular potentials
$V(d) = d^{-p}$ (where $d$ denotes the intermolecular distance and $p > 1$)
give: $I(g, \chi) = g^{- 4/p} \tilde{I}(\chi)$, so that in the
particular case $p=4$ (Maxwell molecules) the collision kernel
($g \, I(g, \chi)$) is independent of $g$.

If the collisions between molecules are elastic, namely momentum and
kinetic energy are conserved in each collision, then the center of mass
velocity is preserved
$$
\mathbf{G}=\frac{1}{2} (\mathbf{v}+\mathbf{w})=
\frac{1}{2} (\mathbf{v'}+\mathbf{w'}),
$$
as well as the relative speed
$$
g=\vert \mathbf{v}-\mathbf{w} \vert=\vert \mathbf{v'}-\mathbf{w'} \vert\,.
$$
Consequently, the collision process is reversible, and pre- and post--collision
velocities, corresponding to the pair $\mathbf{v}, \mathbf{w}$, coincide.
The relation between pre- and post-collision variables reads as

\begin{equation} \label{A.2}
\left\{
\begin{array}{rl}
\mathbf{v'}=\mathbf{G}+\frac{1}{2} g \, \hat{\Omega}' \\
\mathbf{w'}=\mathbf{G}-\frac{1}{2} g \, \hat{\Omega}'
\end{array} \right.
\end{equation}
In other words, post-collision velocities are uniquely determined in terms of
the pre-collision ones, once the direction of the post-collision relative
velocity is given.

(b) Waldmann formulation:

\begin{equation} \label{A.Wald}
Q(f,f)(\mathbf{v})=\int_{\mathbb{R}^3} \int_{\mathbb{R}^3}
\int_{\mathbb{R}^3} W(\mathbf{v}', \mathbf{w}'; \mathbf{v}, \mathbf{w})
[f(\mathbf{v'}) f(\mathbf{w'})-f(\mathbf{v}) f(\mathbf{w}) ] \,
d\mathbf{w} d\mathbf{v}' d\mathbf{w}'.
\end{equation}
In this probabilistic formulation, proposed in \cite{Waldmann}, the kernel
$W(\mathbf{v}', \mathbf{w}'; \mathbf{v}, \mathbf{w})$ represents the
microscopic probability distribution for the collision process
$(\mathbf{v}, \mathbf{w}) \rightarrow (\mathbf{v}', \mathbf{w}')$.
Notice that in (\ref{A.Wald}) the integrals range over all possible values
of the ingoing velocity of the partner molecule $\mathbf{w}$ and of the output
velocity pair $\mathbf{v}', \mathbf{w}'$.
For elastic collisions one has
\begin{equation}
W(\mathbf{v}', \mathbf{w}'; \mathbf{v}, \mathbf{w}) =
\frac{\displaystyle 1}{\displaystyle g} \, I(g, \chi) \,
\delta(\mathbf{G}-\mathbf{G'}) \, \delta(g-g')\,,
\end{equation}
hence, the Waldmann kernel $W$ is symmetric with respect to changes of the
order of particles $\mathbf{v} \leftrightarrow \mathbf{w}$ or of pre- and
post-collision velocities
$(\mathbf{v}, \mathbf{w}) \leftrightarrow (\mathbf{v}', \mathbf{w}')$.

(c) Scattering kernel formulation:

\begin{equation} \label{A.3}
Q(f,f)(\mathbf{v})=\int_{\mathbb{R}^3} \int_{\mathbb{R}^3} \eta(\mathbf{v'},
\mathbf{w'})
\, A(\mathbf{v'}, \mathbf{w'}; \mathbf{v}) \, f(\mathbf{v'}) f(\mathbf{w'})
\, d\mathbf{v'} \, d\mathbf{w'}-
f(\mathbf{v}) \int_{\mathbb{R}^3} \eta(\mathbf{v}, \mathbf{w}) \,
f(\mathbf{w}) \, d\mathbf{w}.
\end{equation}
An analogous probabilistic formulation involves the scattering collision
frequency defined as

\begin{equation} \label{A.4}
\eta(\mathbf{v}, \mathbf{w})=: \int_{\mathbb{R}^3} \int_{\mathbb{R}^3}
W(\mathbf{v}', \mathbf{w}'; \mathbf{v}, \mathbf{w}) \, d\mathbf{v'} \,
d\mathbf{w'} =
\int_{\mathbb{R}^3} \int_{\mathbb{R}^3}
\frac{\displaystyle 1}{\displaystyle g} \, I(g, \chi) \,
\delta(\mathbf{G}-\mathbf{G'}) \, \delta(g-g') \, d\mathbf{v'} \,
d\mathbf{w'}
\end{equation}
and
\begin{equation} \label{A.5}
\eta(\mathbf{v'}, \mathbf{w'}) \, A(\mathbf{v'}, \mathbf{w'}; \mathbf{v})
=:  \int_{\mathbb{R}^3} W(\mathbf{v}', \mathbf{w}'; \mathbf{v}, \mathbf{w}) \,
d\mathbf{w} =
\int_{\mathbb{R}^3} \frac{\displaystyle 1}{\displaystyle g} \, I(g, \chi) \,
\delta(\mathbf{G}-\mathbf{G'}) \, \delta(g-g') \, d\mathbf{w}.
\end{equation}
It can be easily checked that the transition probability
$A(\mathbf{v'}, \mathbf{w'}; \mathbf{v})$ satisfies the properties

\begin{equation} \label{A.8}
A(\mathbf{v'}, \mathbf{w'}; \mathbf{v})=A(\mathbf{w'}, \mathbf{v'};
\mathbf{v})\,,
\end{equation}
\begin{equation} \label{A.9}
\int_{\mathbb{R}^3} A(\mathbf{v'}, \mathbf{w'}; \mathbf{v}) \, d\mathbf{v} =
1\,.
\end{equation}

For Maxwell molecules one has $I(g, \chi) \propto \theta(\chi)/g$, hence
$\eta(\mathbf{v}, \mathbf{w}) = \eta$ (constant) and

\begin{equation} \label{A.7}
A(\mathbf{v'}, \mathbf{w'}; \mathbf{v}) \propto \frac{\theta(\chi^*)}{g'}\,
\delta(v^2-(\mathbf{v'}+\mathbf{w'}) \cdot \mathbf{v}+\mathbf{v'} \cdot
\mathbf{w'})
\end{equation}
where

$$\cos (\chi^*)=\frac{2 \mathbf{v} \cdot (\mathbf{v'}-\mathbf{w'})-v'^2+w'^2}
{\vert \mathbf{v'}-\mathbf{w'} \vert^2}.$$
Further details on these computations are available in \cite{BPS}, \cite{SNB},
where it is also proven the equivalence between formulations (a), (b), (c) for
elastic (deterministic) collisions.

The scattering kernel formalism remains valid also with stochastic collisions,
occurring for instance in social or economic problems,
as in simple market economies or in epidemic models.
For stochastic interactions, in which $A(\mathbf{v'}, \mathbf{w'}; \mathbf{v})$
satisfies only the relationships (\ref{A.8}) and (\ref{A.9}),
with $\mathbf{v'}, \mathbf{w'}, \mathbf{v}$ independent variables, one cannot
expect conservation of momentum and energy.
In the literature of kinetic equations for socio-economic sciences,
it is shown the equivalence between the collision-like
Boltzmann equation and Markovian jump-processes, described by transition
probabilities related to the Waldmann formulation of the Boltzmann equation
\cite{Loy-Tosin}.

\section*{Appendix B: Boltzmann operators for a reacting gas mixture}
\label{appB}
\renewcommand{\theequation}{B.\arabic{equation}}
\setcounter{equation}{0}

In this appendix, we present the classical reactive Boltzmann operators,
originally proposed in \cite{Giovangigli,RS}.
Let us consider a mixture of four gas species which, besides all elastic
collisions,
can interact according to the following reversible bimolecular reaction:

\begin{equation} \label{B.1}
1+2 \rightleftarrows 3+4.
\end{equation}
Let us also introduce the symbols $m_i$ and $E_i$, which stand for
particle mass and internal energy of chemical bond of $i$th-species
($i=1, \dots, 4$), respectively.
Notice that the conservation of mass implies the following constraint:
$m_1 + m_2 = m_3 + m_4 =:M$.

We assume that the internal energies $E_i$ are such that

\begin{equation} \label{B.3}
\Delta E= E_3+E_4-E_1-E_2 \geq 0\,.
\end{equation}
Consequently, the direct reaction $1+2 \rightarrow 3+4$ is endothermic, thus
providing an increase of chemical energy and a decrease of kinetic energy
(total energy, i.e. kinetic plus internal, is preserved); on the other hand,
the reverse reaction $3+4 \rightarrow 1+2$ is exothermic, giving a decrease of
the internal energy.

In such reactive frame, the Boltzmann equation for the distribution $f_i$ reads
as

\begin{equation} \label{B.4}
\frac{\partial f_i}{\partial t}(t, \mathbf{v})=\sum_{j=1}^{4} Q^{ij}+
\overline{Q}^i \; \; \; \; \; \; \; i=1, \ldots, 4,
\end{equation}
where $Q^{ij}$ is the elastic binary operator for collisions involving a pair
of species $(i,j)$, while $\overline{Q}^i$ is the chemical collision operator,
describing the effects on species $i$ due to the reaction (\ref{B.1}).
When a pair of particles $(i,j)$, with molecular velocities
$(\mathbf{v}, \mathbf{w})$, collides, yielding a pair of molecules $(h,k)$
with velocities
$\mathbf{v'}$ and $\mathbf{w'}$, respectively, the differential scattering
cross section in the collision kernel is labeled by $I_{ij}^{hk}$.
In elastic collisions one has $(h,k)=(i,j)$, while in the chemical operator
$(i,j,h,k) \in \{ (1,2,3,4), (2,1,4,3), (3,4,1,2), (4,3,2,1) \}$.

In Eq. (\ref{B.4}), the elastic operator $Q^{ij}$ reads as

\begin{equation} \label{B.5}
Q^{ij}=\int_{\mathbb{R}^3} d\mathbf{w} \int_{\mathbb{S}^2} g \,
I_{ij}^{ij} (g, \hat{\Omega} \cdot
\hat{\Omega}') \, [f_i(\mathbf{v'}) \, f_j(\mathbf{w'})-
f_i(\mathbf{v}) \, f_j(\mathbf{w})] \, d\hat{\Omega}'
\end{equation}
where, taking into account momentum and energy conservation, the
post-collision velocities can be written as

\begin{equation} \label{B.6}
\left\{
\begin{array}{rl}
\mathbf{v'}=\frac{\displaystyle [m_i \mathbf{v}+m_j \mathbf{w}
+m_j g \hat{\Omega}']}{\displaystyle (m_i+m_j)}  \\
\mathbf{w'}=\frac{\displaystyle [m_i \mathbf{v}+m_j \mathbf{w}
-m_i g \hat{\Omega}']}{\displaystyle (m_i+m_j)}
\end{array} \right.
\end{equation}

Chemical operators $\overline{Q}^i$ involve also a suitable Heaviside
function $H$, which accounts for the fact that the endothermic
reaction cannot occur if the kinetic energy of the ingoing molecules is not
enough, namely if the relative speed does not overcome a threshold
(activation energy) depending on the internal energy gap $\Delta\!E$
and on the reduced masses
$$\mu_{ij}=\frac{\displaystyle m_i m_j}{\displaystyle (m_i+m_j)}\,.$$
If the following microreversibility of reactions is invoked,
which relates the cross
section of the direct reaction to that of the reverse one:

\begin{equation} \label{micro}
\big( \mu_{12} \big)^2 g^2\, I_{12}^{34}(g,\hat{\Omega} \cdot \hat{\Omega}') =
\big( \mu_{34} \big)^2 (g')^2\, I_{34}^{12}(g',\hat{\Omega} \cdot
\hat{\Omega}') \qquad \text{for} \qquad g^2 > \frac{\displaystyle 2 \Delta E}
{\displaystyle \mu_{12}}\,,
\end{equation}
then, the chemical collision term in Eq. (\ref{B.4}) for species $1$ is
given by

\begin{equation} \label{B.7}
\overline{Q}^1=\int_{\mathbb{R}^3} \int_{\mathbb{S}^2} H(g^2-
\frac{\displaystyle 2 \Delta E}{\displaystyle \mu_{12}}) \, g \,
I_{12}^{34} (g, \hat{\Omega} \cdot \hat{\Omega}') \bigg[ \bigg(
\frac{\displaystyle \mu_{12}}{\displaystyle \mu_{34}} \bigg)^3 \,
f_3(\mathbf{v'}) f_4 (\mathbf{w'})-f_1(\mathbf{v}) f_2(\mathbf{w}) \bigg]
\, d\mathbf{w} \, d\hat{\Omega}'
\end{equation}
where post-collision velocities take the form
\begin{equation}
\left\{
\begin{array}{rl}
\mathbf{v'}= \frac{1}{M} \Big( m_1 \mathbf{v} + m_2 \mathbf{w} + m_4\, g'\,
\hat{\Omega}' \Big)  \\
\mathbf{w'}= \frac{1}{M} \Big( m_1 \mathbf{v} + m_2 \mathbf{w} - m_3\, g'\,
\hat{\Omega}' \Big)
\end{array} \right.
\qquad \text{with} \qquad g' = \left[ \frac{\mu_{12}}{\mu_{34}}
\left( g^2 - \frac{2\, \Delta\!E}{\mu_{12}} \right) \right]^{1/2}
\end{equation}

For species $3$, the computation can be carried out in exactly the same way,
with the
only important difference that the reverse exothermic reaction (causing a loss
term for species $3$) may occur
for any relative speed of the colliding particles; thus, in the kernel
there is no need of a Heaviside function, and the operator turns out to be

\begin{equation} \label{B.8}
\overline{Q}^3=\int_{\mathbb{R}^3} \int_{\mathbb{S}^2}
 g \,
I_{34}^{12} (g, \hat{\Omega} \cdot \hat{\Omega}') \bigg[ \bigg(
\frac{\displaystyle \mu_{34}}{\displaystyle \mu_{12}} \bigg)^3 \,
f_1(\mathbf{v'}) f_2 (\mathbf{w'})-f_3(\mathbf{v}) f_4(\mathbf{w}) \bigg]
\, d\mathbf{w} \, d\hat{\Omega}'
\end{equation}
where now post-collision velocities are given by

\begin{equation}
\left\{
\begin{array}{rl}
\mathbf{v'}= \frac{1}{M} \Big( m_3 \mathbf{v} + m_4 \mathbf{w} + m_2\, g'\,
\hat{\Omega}' \Big)  \\
\mathbf{w'}= \frac{1}{M} \Big( m_3 \mathbf{v} + m_4 \mathbf{w} - m_1\, g'\,
\hat{\Omega}' \Big)
\end{array} \right.
\qquad \text{with} \qquad g' = \left[ \frac{\mu_{34}}{\mu_{12}}
\left( g^2 + \frac{2\, \Delta\!E}{\mu_{34}} \right) \right]^{1/2}
\end{equation}
The collision terms for species $2$ and $4$ are analogously calculated by
suitable permutations of indices in $\overline{Q}^1$ and $\overline{Q}^3$,
respectively.

Concerning Maxwell molecules, if for the direct reaction we assume a constant
value of
$$
\nu_{12}^{34} = \int_{\mathbb{S}^2} I_{12}^{34} (g, \hat{\Omega} \cdot
\hat{\Omega}')  d\hat{\Omega}'\,,
$$
then, taking into account microreversibility (\ref{micro}), for the cross
section of the reverse reaction one gets
$$
\int_{\mathbb{S}^2} I_{34}^{12} (g, \hat{\Omega} \cdot \hat{\Omega}')
d\hat{\Omega}' = \left( \frac{\mu_{12}}{\mu_{34}} \right)^{3/2}
\left( 1 + \frac{2 \Delta\!E}{\mu_{34}\, g^2} \right) \nu_{12}^{34}\,.
$$

\section*{Acknowledgements}

M.B. and S.L. are supported by GNFM of INdAM, Italy. 
Moreover, M.B. thanks the support by the Unversity of Parma through the 
project {\it Collective and self-organized dynamics: kinetic and network 
approaches} 
(Bando di Ateneo 2022 per la ricerca - PNR - PNRR - NextGenerationEU), and 
by the Italian Ministry MUR through the projects 
{\it Integrated Mathematical Approaches to Socio-Epidemiological Dynamics} 
(PRIN 2020JLWP23) and {\it Mathematical Modelling for a Sustainable Circular 
Economy in Ecosystems} (PRIN P2022PSMT7).

\section*{Declarations}

\par\noindent
\textbf{Data Availability}
\par\noindent
Tha data that support the findings of this study are available from the
corresponding author.

\par\noindent
\textbf{Conflict of interest}
\par\noindent
The authors declare that they have no conflict of interests.

%\clearpage
%\newpage

\end{document}